\documentclass{jssc}

\newtheorem{condition}{Condition}
\newcommand{\col}{\text{col}\,}
\newcommand{\diag}{\text{diag}\,}
\newcommand{\N}{\mathcal{N}}
\newcommand{\T}{\mathcal{T}}

\usepackage{algorithm}%
\usepackage{algorithmicx}%
\usepackage{algpseudocode}%
\usepackage{booktabs}
\usepackage{xcolor}

\def\textsubscript#1%
{$_{\text{#1}}$}

\def\cdd{\mbox{\boldmath$\cdot$}~}

\input{amssym.def}
\include{graphix}

\makeatletter
\def\@oddfoot{\hfill}
\newcount\shumeicount
\def\setshumei#1#2#3{%
  \shumeicount=\count0
  \def\@oddhead{%
    \raise-5pt\hbox to0pt{\vrule width\hsize height 0pt depth 0.4pt\hss}\relax
    \ifnum \shumeicount=\count0
      \raise-7pt\hbox to0pt{\vrule width\hsize height 0pt depth 0.4pt\hss}\relax
      #1
    \else
      \ifodd\count0
        #2
      \else
        #3
       \fi
     \fi
  }%
}
\makeatother
\makeatletter
\def\@oddfoot{\hfill}
\newcount\shujiaocount
\def\setshujiao{%
  \shujiaocount=\count0
  \def\@oddfoot{%
      \ifodd\count0
      \else
      \fi
  }%
}
\makeatother
\def\title#1#2#3#4{{
  \vspace*{0.3cm}
  \begin{flushleft} \Large\bf #1\end{flushleft}
  \vspace*{-0.2cm}
      \begin{flushleft}
      \bf #2
      \end{flushleft}
      \footnotetext{\hspace{-6mm} #3\\ #4}}}

\def\dshm#1#2#3#4
{\setshumei{J Syst Sci Complex (#1) #2:
{\thepage--\pageref{LastPage}}\hfill}
            {\hfill {\small #3}\hfill\hbox to0pt{\hss\thepage}}
            {\hbox to0pt{\thepage\hss}\hfill {\small #4}\hfill
            }
            \setshujiao}
\def\drd#1#2
{{\vskip 1cm\small \begin{flushleft}
 #1 \\
 #2\\
\copyright The Editorial Office of JSSC \&  Springer-Verlag GmbH
Germany 2021
\end{flushleft}}}

\def\hat{\widehat}
\def\tilde{\widetilde}
\def\bar{\overline}
\def\epsilon{\varepsilon}

\begin{document}

\title{On Finite-Horizon Approximation of an Infinite-Horizon Feedback Nash Equilibrium in Discrete-Time LQ Games$^*$}
{\uppercase{Huang} Shengyuan \cdd \uppercase{Yang}
Xiaoguang \cdd \uppercase{Mu} Yifen \cdd \uppercase{Mei} Wenjun}
{\uppercase{Huang} Shengyuan \cdd \uppercase{Mu} Yifen\\
Academy of Mathematics and Systems Science, Chinese Academy of Sciences, Beijing 100190, China; University of Chinese Academy of Sciences, Beijing 100190, China.  Email: huangshengyuan22@mails.ucas.ac.cn; mu@amss.ac.cn.\\  
\uppercase{Yang} Xiaoguang   \\
School of Economics and Management, China University of Petroleum-Beijing, Beijing 102249, China; Academy of Mathematics and Systems Science, Chinese Academy of Sciences, Beijing 100190, China; University of Chinese Academy of Sciences, Beijing 100190, China.  Email: xgyang@iss.ac.cn.\\
\uppercase{Mei} Wenjun (Corresponding author) \\
Department of Mechanics
and Engineering Science, Peking University, Beijing 100871, China.  Email: mei@pku.edu.cn. \\
   } 
{Wenjun Mei is supported by the National Natural Science Foundation of China (Grant No. 72531001, No. 72201008). Xiaoguang Yang is supported by the National Natural Science Foundation of China (Grant No. 72192804).}



\dshm{2026}{09}{On Finite-Horizon Approximation of an Infinite-Horizon
FNE in Discrete-Time LQ Games}{\uppercase{Huang 
Shengyuan} $\cdd$ \uppercase{Yang Xiaoguang} $\cdd$
\uppercase{Mu Yifen} \cdd
\uppercase{Mei Wenjun}}
\thispagestyle{empty}
\Abstract{Computing feedback Nash equilibria (FNEs) in infinite-horizon discrete-time linear-quadratic (LQ) dynamic games remains computationally challenging. Inspired by model predictive control (MPC) in single-agent optimal control, we address this challenge with a finite-horizon strategy for approximating one such FNE. 
 The finite-horizon strategy is as follows. Each player $i$ has an individual prediction horizon $T^i$. At each stage, player $i$ envisions an auxiliary $T^i$-stage game, computes its unique FNE, and implements only the first-stage control.  
Our main results are as follows.
First, we give parameter conditions that guarantee geometric convergence of the coupled Riccati iteration to a stabilizing solution. 
Second, under these conditions, the finite-horizon strategies stabilize the system, and each player's total cost converges to the limiting FNE cost as all prediction horizons tend to infinity.
Third, we derive an explicit upper bound on this cost gap that decreases geometrically with the shortest prediction horizon. This bound tells us how long the prediction horizons need to be for a given accuracy. 
The strategy is tractable and implementable, as it avoids directly solving the coupled algebraic Riccati equations of the infinite-horizon game.}      


\section{Introduction}
Dynamic games provide a framework for modeling multi-stage
competition and cooperation and have been widely studied in
robotics \cite{robotics_game2025,robotics_attack2022}, control
theory \cite{1998basarNoncooperativeGame,TACgame2024,siam2024,automatica_game2025},
economics \cite{econometrica2022,aer2017}, and biology
\cite{Science_2004}. 
A fundamental solution concept is the feedback Nash equilibrium
(FNE) \cite{1998basarNoncooperativeGame,engwerda2005lq,siam2024,Liu_Guo_2024,tac_fne_2022}.
Under an FNE, each player chooses actions based on the current
state, and no player can reduce its total cost by changing its
strategy while the other players' strategies remain fixed.
However, in infinite-horizon linear-quadratic (LQ) dynamic games,
the coupled algebraic Riccati equations associated with FNEs are
generally difficult to solve because of their coupling and
nonlinearity \cite{1998basarNoncooperativeGame,engwerda2005lq}.
This difficulty has motivated research on iterative methods
\cite{TACgame2024,letters_iterative_2020,Automatica_iterative_2023,iteration_2024}
and approximation methods
\cite{opt_2024,approximate_siopt_2023,sinica_2014,IFAC_Nortmann_Mylvaganam_2023}.

While continuous-time differential games have been extensively
studied, discrete-time games receive comparatively less attention.
For infinite-horizon linear quadratic discrete-time dynamic games,
Nortmann et al.~\cite{TACgame2024} propose four algorithms to
compute FNEs via policy or value iteration and discuss conditions
under which an FNE is a locally asymptotically stable (LAS)
equilibrium of the iteration. However, verifying the LAS
conditions is technically demanding: it requires constructing
a mapping from the vectorized coupled Riccati equations,
linearizing this mapping around an FNE, and checking that the
matrix obtained from the resulting Jacobians is Schur stable.
Moreover, evaluating these Jacobians requires knowledge of the
FNE being sought, making the conditions difficult to verify
a priori.

To alleviate computational difficulty, Nortmann et
al.~\cite{IFAC_Nortmann_Mylvaganam_2023} introduce the notion of
an approximate FNE, namely the $\epsilon_{\alpha,\,\beta}$-Nash
equilibrium, which generalizes the $\epsilon_\alpha$-Nash
equilibrium in~\cite{tac_approximate_2014}.
In~\cite{IFAC_Nortmann_Mylvaganam_2023}, the coupled discrete
Riccati equations associated with FNEs are reformulated as a
non-convex semidefinite program (SDP), allowing the approximate
equilibrium to be computed more easily than the exact FNE.  However, ~\cite{IFAC_Nortmann_Mylvaganam_2023} does not establish convergence of
zero-terminal-cost finite-horizon FNEs as the prediction
horizon increases.

These computational challenges also motivate the study of receding-horizon methods, which apply the model predictive control (MPC) principle to dynamic games. At each stage, the current state serves as the initial condition of a finite-horizon problem, and only the corresponding first-stage actions are implemented. The prediction window then moves forward by one stage. Our finite-horizon strategy follows this principle: each player $i$ chooses a fixed horizon $T^i$, solves an auxiliary game involving all players over that horizon with zero terminal cost, and implements its own first-stage FNE action.

For feedback equilibria in discrete-time LQ games, Benenati and
Grammatico~\cite{BenenatiGrammatico2026} develop a receding-horizon
construction using a known infinite-horizon FNE. Its cost
matrices provide the terminal weights, and its gains are used
to predict the opponents' future actions. Our auxiliary games
instead use zero terminal cost and require no prior knowledge
of the limiting FNE.

With zero terminal cost, one must establish that the
finite-horizon Riccati recursion is well defined and converges.
Salizzoni et al.~\cite{SalizzoniHallKamgarpour2025} establish
the relation between its fixed points and stationary
infinite-horizon FNEs, but study convergence numerically.
We give explicit parameter conditions under which the recursion
is well defined and converges geometrically, together with
explicit error bounds for the feedback and cost matrices.

A further question concerns the stability and costs of the
implemented strategies. Hall et al.~\cite{Hall2026Stability}
give stability conditions for constrained discrete-time games
with linear dynamics and no terminal costs. For demand-side
management, Hall et al.~\cite{Hall2022Demand} also bound the
state's long-run distance from a steady point. This bound tends
to zero as the horizon grows. Both methods compute open-loop
equilibrium action sequences over a common horizon and
recompute them from the current state at each stage. Their
implementation therefore also uses state feedback. Within each
prediction window, however, we compute feedback equilibria,
which specify how future actions depend on future states.
We allow different horizons and bound each player's actual
infinite-horizon cost relative to its cost under the limiting FNE.

We consider unconstrained LQ games with known linear
input/output/state (i/o/s) dynamics, observed states, and
player-specific discount factors. Costs are quadratic in output
tracking errors and control inputs~\cite{DEEPC2019ECC,lqbook_track2007}.
This form is useful in our related work on games with unknown
dynamics and unobserved states~\cite{huang2025unknown}, since
stage costs can be evaluated directly from input-output data. 
Appendix~\ref{app:ios_equivalence} gives an equivalent input/state
(i/s) formulation and explains the relations between our model
and standard i/s LQ games. Our infinite-horizon analysis concerns zero references and
stationary linear feedback equilibria, with unilateral deviations
restricted to those preserving closed-loop stability.

As preliminaries, we first recall standard finite-horizon LQ game results in the i/o/s form used in this paper. We then establish three main results for the infinite-horizon game with zero reference.
First, we give explicit sufficient conditions on the model parameters under which the coupled Riccati iteration is well defined and converges geometrically to a stabilizing solution. The limit determines the FNE selected by zero terminal cost, even when other FNEs exist.
Second, under suitable conditions, we show that the finite-horizon strategies with possibly different individual prediction horizons stabilize the system when the shortest horizon is sufficiently large. Each player's actual infinite-horizon cost converges to its cost under the limiting FNE as the shortest horizon tends to infinity.
Third, under the explicit sufficient parameter conditions above, stability holds for every choice of positive individual prediction horizons, and the gap between each player's actual cost and its limiting FNE cost admits an explicit bound that decreases geometrically with the shortest horizon. The bound depends only on the model parameters and initial state, so players can compute a sufficient common lower bound on their horizons before play for a given cost tolerance.
Finally, we provide two non-scalar numerical examples with nonzero constant references. They illustrate the effects of individual prediction horizons and player-specific discount factors beyond the zero-reference setting covered by our theory.

The remainder of this paper is organized as follows.
Section~\ref{sec:known-dynamics Game} recalls standard
finite-horizon LQ game results in i/o/s form.
Section~\ref{sec:Infinite-Horizon Discrete-Time LQ} presents the
finite-horizon strategy and our main results on convergence,
stability, and error bounds.
Section~\ref{sec:simulation} presents two non-scalar numerical examples. Section~\ref{sec:conclusions} concludes the paper.
Appendix~\ref{app:ios_equivalence} presents an equivalent i/s formulation and compares our model with standard classes of i/s LQ games, 
Appendices~\ref{app:lm:Uniqueness_FN}--\ref{app:Proof of Theorem}
contain the proofs, and Appendix~\ref{app:simulation-fnes} proves
the existence of infinitely many stabilizing affine FNEs
in the second example.

\section{Preliminaries: Standard Finite-Horizon LQ Game in i/o/s Form}
\label{sec:known-dynamics Game}

This section recalls standard finite-horizon LQ dynamic game results from two classical textbooks~\cite{1998basarNoncooperativeGame, engwerda2005lq}, stated in the i/o/s form used in this paper. We adopt the i/o/s dynamics and objective functions of this form to model games involving output-trajectory tracking, where tracking errors are naturally specified in terms of outputs while the strategic interactions are governed by states and inputs.  The results recalled in this section serve as a necessary basis for our main results.

\paragraph{Basic Notations}
Let $\mathbb{N}$ and $\mathbb{R}$ denote the sets of natural and real numbers, respectively. 
Let $\mathbf{0}$ denote the zero vector or matrix of appropriate dimension. For a symmetric matrix $M$, we write $M\succeq 0$ (resp., $M\succ 0$) to indicate that $M$ is positive semi-definite (resp., positive definite). Let \(\|\cdot\|_2\) denote the 2-norm of a matrix. For $t_1,t_2\in\mathbb{N}$ with $t_1\le t_2$ and a sequence $\{a_t\}_{t\in\mathbb{N}}$, define $a_{t_1:t_2}=(a_{t_1},\dots,a_{t_2})$. For matrices $A_1,\dots,A_m$, let $\diag(A_1,\dots,A_m)$ denote the block-diagonal matrix with diagonal blocks $A_1,\dots,A_m$. If these matrices have the same number of columns, define
\begin{align*}
    \col(A_1,\dots,A_m):=[A_1^\top,\dots,A_m^\top]^\top .
\end{align*}

\paragraph{Game Setup}
Let $\N=\{1,\dots,N\}$ and $\T=\{1,\dots,T\}$ denote the sets of players and time stages, respectively. Each player $i$ controls the input $u_t^i \in \mathbb{R}^{m_i}$ for any $t \in \mathcal{T}$. Define the stacked control vector $u_t = \operatorname{col}(u_t^1,\ldots,u_t^N)$ and the
block matrices $B=[B^1, \dots, B^N]$ and $D=[D^1, \dots, D^N]$. Consider the discrete-time linear system
\begin{equation}
\label{eq:isosystem}
    \begin{cases}
        x_{t+1} = Ax_t + \sum\limits_{i=1}^N B^i u^i_t = Ax_t + B u_t,\\[2mm]
        y_t = Cx_t + \sum\limits_{i=1}^N D^i u^i_t = Cx_t + Du_t,
    \end{cases}
\end{equation}
where $x_t\in\mathbb{R}^n$ and $y_t\in\mathbb{R}^p$ denote the state and output, respectively.   Player $i$ aims to minimize the following finite-horizon cost:
\begin{equation}
\label{eq:costfun}
    \begin{aligned}
    J^i(x_1,u_{1:T})
    =\frac{1}{2}\sum_{t=1}^{T} 
    \Bigg[
     (y_t-l_t^i)^\top Q^i (y_t-l_t^i)  +\sum_{j=1}^N (u_t^j)^\top R^{ij} u_t^j
    \Bigg]\,(\delta^i)^{t-1},
    \end{aligned}
\end{equation}
where $l_{1:T}^i$ is the reference output trajectory of player $i$, $\delta^i\in(0,1]$ is the individual discount factor, and  
$Q^i\succeq 0$, $R^{ii}\succ 0$, $R^{ij}\succeq 0$ hold for any $i,j \in \N$. 

The above setting defines an $N$-player $T$-stage dynamic game, referred to as a \emph{finite-horizon LQ game with input/output/state (i/o/s) dynamics}. This game can be rewritten as an i/s LQ game with time-invariant dynamics and time-varying costs. Appendix~\ref{app:ios_equivalence} gives the details and discusses the relations between our model and standard i/s LQ games. In particular, our model strictly contains the class of time-invariant i/s LQ games whose stage costs contain no cross, linear, or constant terms.

We assume that all players have complete knowledge of the model parameters 
$A,B,C,D,Q^i,$ $R^{ij},l_t^i,\delta^i$ for any $i, j \in \N$ and $t \in \mathcal{T}$, and that all players observe the system state $x_t$ at each stage $t \in \T$. This information structure is standard in dynamic games~\cite{1998basarNoncooperativeGame,engwerda2005lq,automatica_game2025}.

For any $i\in\N$, a feedback strategy is a mapping $\gamma_t^i:\mathbb{R}^n\to\mathbb{R}^{m_i}$ such that $u_t^i=\gamma_t^i(x_t)$. Let $\gamma_t=(\gamma_t^1,\dots,\gamma_t^N)$ denote the profile of players' strategies at stage $t$ for any $t \in \T$. The strategy profiles from time $1$ to time $T$ are denoted by $\gamma_{1:T}=(\gamma_1,\dots,\gamma_T)$. Given the initial state $x_1$ and $\gamma_{1:T}$, the trajectories $\{x_t,y_t,u_t\}_{t=1}^T$ are uniquely determined by dynamics~\eqref{eq:isosystem}. Accordingly, we use the shorthand $J^i(x_1,\gamma_{1:T})$ to denote the induced cost. For convenience, we use $\gamma_t^{(-i)}$ to denote the strategies of all players except $i$ at stage $t$. We now recall the notion of feedback Nash equilibrium.

\begin{definition}[Feedback Nash Equilibrium (FNE), {\citep[Definition~3.22]{1998basarNoncooperativeGame}}]
\label{de:FNE_finite}
A strategy profile $(\gamma_{1:T}^{1*},\dots,\gamma_{1:T}^{N*})$ is a feedback Nash equilibrium if, for any initial state $x_1\in\mathbb{R}^n$, any $t\in\T$, and any $i\in\N$,
\begin{align*}
J^i(x_1,\gamma_{1:t-1},\gamma_{t:T}^{i*},\gamma_{t:T}^{(-i)*})
\le
J^i(x_1,\gamma_{1:t-1},\gamma_{t:T}^{i},\gamma_{t:T}^{(-i)*}),
\end{align*}
holds for all admissible strategies
$\gamma_1,\dots,\gamma_{t-1},\gamma_t^i,\dots,\gamma_T^i$. 
\end{definition}

\paragraph{Standard Results of Feedback Nash Equilibrium}
The following lemma presents a necessary and sufficient condition for the existence of a unique FNE of discrete-time  finite-horizon LQ  games.
\begin{lemma}[{\citep[Corollary~6.1]{1998basarNoncooperativeGame}}]
\label{lm:suffi nece}
The $N$-person $T$-stage finite-horizon LQ game ~\eqref{eq:isosystem},~\eqref{eq:costfun} has a unique FNE $(\gamma_{1:T}^{1*},\dots, \gamma_{1:T}^{N*})$ if and only if the following discrete coupled Riccati difference equations admit a unique solution \\
$\{P^{i*}_t, S^{i*}_t, w^{i*}_t, K^{i*}_t, L^{i*}_t\,|\,i\in \N, t\in \T\}$:
\begin{align}
    \label{eq:K} 
    & \delta^i (B^i)^\top P_{t+1}^i F_t  +(D^i)^\top Q^i G_t + R^{ii} K^i_t = \mathbf{0}, \\
    \label{eq:L}  
    & \begin{aligned}
        \delta^i (B^i)^\top (P_{t+1}^i I_t & +  S^i_{t+1})+ R^{ii} L^i_t +(D^i)^\top Q^i M_t^i = \mathbf{0}, 
    \end{aligned}\\
    \label{eq:P} & 
    \begin{aligned}
        P^i_t  =  & G_t^\top Q^i G_t+\delta^i F_t^\top P^i_{t+1}F_t+ 
\sum\limits_{j\in N}(K^j_t)^\top R^{ij} (K^j_t),
    \end{aligned} \\
    \label{eq:Q} &
    \begin{aligned}
        S^i_t = & G_t^\top Q^i M_t^i + \delta^i F_t^\top P^i_{t+1}I_t + \delta^i F_t^\top S^i_{t+1} + \sum\limits_{j\in N}(K^j_t)^\top R^{ij} L^j_t, 
    \end{aligned} \\
    \label{eq:w} & 
    \begin{aligned}
        w^i_t = & \frac{1}{2}(M_t^i)^\top Q^i M_t^i  + \delta^i \frac{1}{2}(I_t)^\top P^i_{t+1} I_t   + \delta^i (I_t)^\top S^i_{t+1}+ \delta^i w^i_{t+1} + \frac{1}{2}\sum\limits_{j\in N}(L^j_t)^\top R^{ij} (L^j_t) ,
    \end{aligned} 
\end{align}
with $P^i_{T+1}=\mathbf{0},  S^i_{T+1}=\mathbf{0}, w^i_{T+1}=0$ for any $i \in \N$. Here $F_t = A+\sum\limits_{j\in N}B^j K^j_t$, $G_t = C+\sum\limits_{j\in N}D^j K^j_t$, $I_t = \sum\limits_{j\in N}B^j L^j_t$ and $M_t^i = \sum\limits_{j\in N}D^j L^j_t - l^i_t$, for any $i \in \N$ and any $t\in \T$. Given the solution to the above equations, the FNE strategy and each player's total cost are given by 
 \begin{equation*}
    \begin{aligned}
\left\{\begin{aligned}
&u_t^{i*}=\gamma^{i*}_t(x_t) = K^{i*}_t x_t + L^{i*}_t,\\
&J^{i}(x_1,u^*_{1:T}) = \frac{1}{2} x_1^\top P_1^{i*} x_1 + (S^{i*}_1)^\top x_1 + w^{i*}_1,
\end{aligned}\right.
\end{aligned}
\end{equation*}
for any $t\in \T$ and any $i\in \N$.
\end{lemma}

This lemma follows from standard backward induction and the Bellman equations~\citep[Corollary~6.1]{1998basarNoncooperativeGame}. In what follows, we recall Remark 6.5 in~\cite{1998basarNoncooperativeGame} in this paper's setting, which provides a sufficient condition for the uniqueness of the FNE. Based on this condition, one can straightforwardly construct a computationally convenient algorithm for the unique FNE. To this end, we present the block matrix formulation of the discrete coupled Riccati difference equations~\eqref{eq:K}$\sim$\eqref{eq:w}.

For any $t \in \T$, given $P_{t+1}^i$ for all $i \in \N$, equation~\eqref{eq:K} becomes a set of linear equations in $K_t^i$, which can be equivalently written as 
\begin{equation}
H(P_{t+1}) K_{t} = g(P_{t+1}),
\label{eq:K-Merge}
\end{equation}
where \(P_{t+1}= \col(P^1_{t+1},\dots,P^N_{t+1})\) and $K_{t}= \col(K^1_{t},\dots,K^N_{t})$ for any $t \in \T$.  The matrices \(H(P_{t+1})\) and \(g(P_{t+1})\) are affine functions of 
 \(P_{t+1}\), and their detailed structures are given by the following equations, 
\begin{equation}
\label{eq:linear_H}
\begin{aligned}
H(P_{t+1}) = & \diag(D^1,\dots, D^N)^{\top}\col(Q^1,\dots,Q^N)D + \diag( R^{11},\dots, R^{NN} ) \\
&+ \diag( \delta^1 B^1,\dots,\delta^N B^N )^{\top}P_{t+1}B
\end{aligned}
\end{equation}

\begin{equation}
\label{eq:linear_g}
\begin{aligned}
g(P_{t+1}) = & -\diag(D^1,\dots, D^N)^{\top}\col(Q^1,\dots,Q^N)C\\
& - \diag( \delta^1 B^1,\dots,\delta^N B^N )^{\top}P_{t+1} A.
\end{aligned}
\end{equation}
For any $t \in \T$, given \(P_{t+1}^i, S_{t+1}^i\) for all $i \in \N$, equation~\eqref{eq:L} becomes a set of linear equations in \(L^i_{t}\), written as 
\begin{align*}
    \tilde{H}(P_{t+1}) L_{t} = \tilde{g}(S_{t+1}),
\end{align*}
where $S_{t+1}= \col(S^1_{t+1},\dots,$ $ S^N_{t+1})$ and $L_{t}= \col(L^1_{t},\dots,L^N_{t})$ for any $t \in \T.$
It can be verified that \(\tilde{H}(P_{t+1})=H(P_{t+1})\), which leads to
\begin{equation}
\label{eq:L-Merge}
H(P_{t+1}) L_{t} = \tilde{g}(S_{t+1}),
\end{equation}
where \(\tilde{g}(S_{t+1})\) is the following affine function of \(S_{t+1}\):
\begin{equation}
\label{eq:linear_g'}
\begin{aligned}
\tilde{g}(S_{t+1})
= {} & \diag\!\Big((D^1)^\top Q^1,\dots,(D^N)^\top Q^N\Big)
\col(l_t^1,\dots,l_t^N) \\
&-\diag(\delta^1 B^1,\dots,\delta^N B^N)^\top S_{t+1}.
\end{aligned}
\end{equation}

Based on equations~\eqref{eq:K-Merge} and~\eqref{eq:L-Merge}, as well as the simple update formulas given by equations~\eqref{eq:P}$\sim$\eqref{eq:w}, we recall the following standard sufficient condition for the existence of a unique FNE.

\begin{lemma}[{\citep[Corollary~6.1 and Remark~6.5]{1998basarNoncooperativeGame}}]

\label{lm:Uniqueness_FNE} 
Consider the finite-horizon LQ game~\eqref{eq:isosystem},~\eqref{eq:costfun}. Construct $\{P_t\}_{1\le t\le T+1}$ iteratively via equations~\eqref{eq:K}$\sim$\eqref{eq:w}, with $P^i_{T+1}= \mathbf 0,\, S^i_{T+1}=\mathbf 0$, and $w^i_{T+1}=0$, for all $i\in \N$. If $|H(P_{t+1})| \neq 0$ for any $t\in \T$, then this game has a unique FNE.
\end{lemma}

The proof of Lemma~\ref{lm:Uniqueness_FNE} is given in Appendix~\ref{app:lm:Uniqueness_FN}, which straightforwardly leads to the following simple algorithm for computing the unique FNE of the finite-horizon LQ game. If $H(P_{t+1})$ is invertible for all $t\in\T$, which can be verified during the iteration, the algorithm computes the FNE by solving $T$ linear systems. If the loop completes all $T$ steps and reaches $t=0$,
the algorithm returns the unique FNE. If it stops with
$t\ge1$, then $H(P_{t+1})$ is singular, and the algorithm
reports failure without returning a complete strategy profile.

\begin{algorithm}
\caption{Backward Algorithm for $T$-Stage Games}
\label{alg:Algorithm1}
\begin{algorithmic}[1]
\State \textbf{Input:} $T,A, B, C, D, x_1,$ $Q^i,  \delta^i, R^{ij},  l^i_t$ \\
\(t=T, P^{i}_{T+1}=\mathbf 0,S^{i}_{T+1}=\mathbf0, w^{i}_{T+1}=0,~t \in \T, ~i,j \in \mathcal{N}\)

\While{$t > 0$ and $|H(P_{t+1})| \neq 0$}
        \State{$K_t \gets H(P_{t+1})^{-1}g(P_{t+1})$}
        \State{$L_t \gets H(P_{t+1})^{-1}\tilde g(S_{t+1})$}
        \State{Obtain $P^{i}_{t},S^{i}_{t}, w^{i}_{t}$ by~\eqref{eq:P},~\eqref{eq:Q},~\eqref{eq:w}}
      \State{ $t \gets t-1$}
\EndWhile
\If{$t>0$}
    \State \Return failure: $H(P_{t+1})$ is singular
\EndIf
\State{\textbf{Output:} $K^{i}_t, L^{i}_t, $ $P^{i}_t, S^{i}_t, w^{i}_t,
i \in \N,t \in \T$}
\end{algorithmic}
\end{algorithm}

\section{Infinite-Horizon Discrete-Time LQ Game and Finite-Horizon Strategy}
\label{sec:Infinite-Horizon Discrete-Time LQ}

This section considers the infinite-horizon discrete-time LQ game with i/o/s dynamics~\eqref{eq:isosystem}, where player $i$'s cost function is given by
\begin{equation}
J^i(x_1,u)=\frac{1}{2} \sum_{t=1}^{+\infty} 
\Bigg[(y_t)^\top Q^i y_t + \sum_{j \in N } (u^j_t)^\top R^{ij} u^j_t\Bigg](\delta^i)^{t-1}. 
\label{eq:costfun_infinite}
\end{equation}
This is the infinite-horizon counterpart of~\eqref{eq:costfun} with $l^i_t= \mathbf 0$ for all $i \in \mathcal N$ and $t \in \mathbb{N}_+$. 
Such purely quadratic infinite-horizon cost functions are standard in the literature; see, e.g.,~\citep{1998basarNoncooperativeGame, engwerda2005lq,TACgame2024}. We consider stationary linear feedback strategies
\[
u_t^i=K^i x_t,\qquad i\in\mathcal N,\quad t\in\mathbb N_+.
\]
Following~\cite{siam2024}, we restrict unilateral deviations to those that keep the closed-loop system Schur stable. Specifically, when all other players use $u_t^j=K^{j*}x_t$, a deviation $u_t^i=\hat K^i x_t$ is admissible if
\[
A+B^i\hat K^i+\sum_{j\ne i}B^jK^{j*}
\]
is Schur stable. A stabilizing strategy profile is an FNE if, for every initial state, no player can reduce its cost through any such deviation. This definition excludes unstable deviations, even when their discounted costs are finite.

We study these equilibria through the coupled Riccati difference equations~\eqref{eq:K} and~\eqref{eq:P} and their infinite-horizon algebraic counterparts. The infinite-horizon game may admit multiple FNEs. Our aim is to approximate one of them.

To facilitate our analysis, we first introduce the notation and
mappings for the coupled Riccati iteration.
For cost block columns
$X=\operatorname{col}(X^1,\ldots,X^N)$ with
$X^i\in\mathbb R^{n\times n}$, define
\[
\|X\|_{\max,P}:=\max_{i\in\mathcal N}\|X^i\|_2.
\]
For feedback block columns
$Y=\operatorname{col}(Y^1,\ldots,Y^N)$ with
$Y^i\in\mathbb R^{m_i\times n}$, define
\[
\|Y\|_{\max,K}:=\max_{i\in\mathcal N}\|Y^i\|_2.
\]
The subscripts $P$ and $K$ distinguish the two prescribed
row partitions. 

For $P=\operatorname{col}(P^1,\ldots,P^N)$ with
$P^i\in\mathbb R^{n\times n}$, whenever $H(P)$ is nonsingular,
define
\[
\mathcal K(P):=H(P)^{-1}g(P),
\]
\[
F(P):=A+B\mathcal K(P),\qquad
G(P):=C+D\mathcal K(P),
\]
and
\[
\Phi^i(P):=G(P)^\top Q^iG(P)
+\delta^iF(P)^\top P^iF(P)
+\sum_{j\in\mathcal N}
\mathcal K^j(P)^\top R^{ij}\mathcal K^j(P).
\]
Here $\mathcal K^i(P)\in\mathbb R^{m_i\times n}$ denotes
the $i$th row block of $\mathcal K(P)$, and
$\Phi^i(P)\in\mathbb R^{n\times n}$ for each $i\in\mathcal N$. 
Define
\[
\Phi(P):=\operatorname{col}(\Phi^1(P),\ldots,\Phi^N(P))
\in\mathbb R^{Nn\times n}.
\] 
The maps $\mathcal K$ and $\Phi$ follow from the Riccati
equations~\eqref{eq:K} and~\eqref{eq:P}, respectively.
Starting from $P^{[0]}=\mathbf 0$, consider the Riccati iteration
\[
K^{[k]}=\mathcal K(P^{[k]}),\qquad
P^{[k+1]}=\Phi(P^{[k]}),\qquad k=0,1,\ldots.
\]
We introduce the following condition on the Riccati iteration and
its limit.

\begin{condition}
\label{assup:converge}
\begin{enumerate}
\item[$(i)$] Invertibility: $H(P^{[k]})$ is nonsingular for every $k\ge0$.

\item[$(ii)$] Convergence: The limits
\[
\lim_{k\to\infty}P^{[k]}=P^*,\qquad
\lim_{k\to\infty}K^{[k]}=K^*
\]
exist and are finite.

\item[$(iii)$] Stability: $\|A+BK^*\|_2<1$.
\end{enumerate}
\end{condition}

In single-agent LQ control, uniqueness of the finite-horizon
optimal control follows from the standard definiteness conditions
$R\succ\mathbf0$ and $Q\succeq\mathbf0$; see
\cite[Proposition~5.1]{1998basarNoncooperativeGame}.
In multi-agent games, however, the corresponding definiteness
conditions alone do not guarantee that the coupled matrix $H(P)$
is nonsingular. We therefore need 
Condition~\ref{assup:converge}(i), which ensures that the Riccati
iteration is well defined and, by Lemma~\ref{lm:Uniqueness_FNE},
guarantees a unique FNE for every finite horizon game. 
Condition~\ref{assup:converge}(ii) requires the cost and feedback
iterates to converge to $P^*$ and $K^*$, respectively. 
Condition~\ref{assup:converge}(iii) ensures stability of the
limiting closed-loop system.

In single-agent LQ control, convergence of the Riccati iteration
is classical under suitable assumptions; see
\cite[Proposition~5.2]{1998basarNoncooperativeGame}.
For coupled Riccati equations, however, establishing convergence
directly from the model parameters is more difficult; see
\cite[Section~6.2.3]{1998basarNoncooperativeGame}.
Condition~\ref{assup:converge} describes properties of the coupled Riccati iteration
and its limit, whose validity is not immediate from the model parameters. 
Therefore, we establish explicit parameter conditions that are
sufficient, though not necessary, to guarantee all three parts of
Condition~\ref{assup:converge}. Under these conditions, we further
derive explicit geometric convergence bounds for both $P^{[k]}$
and $K^{[k]}$.

\subsection{Sufficient Parameter Conditions and Convergence Rates}
\label{subsec: Sufficient parameter conditions and convergence rates}

We first define  
\[
a:=\|A\|_2,\qquad
b:=\sum_{i\in\mathcal N}\|B^i\|_2,\qquad
d:=\sum_{i\in\mathcal N}\|D^i\|_2.
\]
For the cost matrices, define
\[
q:=\max_{i\in\mathcal N}\|Q^i\|_2,\qquad
r:=\min_{i\in\mathcal N}\lambda_{\min}(R^{ii}),\qquad
\overline R:=\max_{i\in\mathcal N}\sum_{j\in\mathcal N}\|R^{ij}\|_2.
\]
Here $\lambda_{\min}$ denotes the smallest eigenvalue, and
the definiteness conditions on the cost matrices imply
$0<r\le\overline R$.
To express the parameter conditions and convergence bounds
compactly, define
\[
\kappa:=\frac{\sqrt{q\overline R}}r,\qquad
s:=q\|C\|_2^2.
\]
For $a<1$, let
\[
M:=\frac{2s}{1-a^2},\qquad
\rho:=a+\frac{1-a}{10},\qquad
\mathcal P_M:=\{P:\mathbf0\preceq P^i\preceq MI_n,\ i\in\mathcal N\}.
\]
We analyze the Riccati map $\Phi$ on $\mathcal P_M$.
The following theorem shows that, under explicit parameter
inequalities, the region $\mathcal P_M$ is invariant under $\Phi$ and
$\rho\in(0,1)$ bounds its contraction factor.

\begin{theorem}[Sufficient parameter conditions]
\label{theo:Sufficient parameter conditions} 
Consider the infinite-horizon game~\eqref{eq:isosystem},
\eqref{eq:costfun_infinite}.
Suppose that
\begin{equation}
a<1,\qquad a+5\kappa\|C\|_2b\le1,\qquad 5\kappa d\le1.
\label{eq:pc-general-conditions}
\end{equation}
Then the following statements hold.
\begin{enumerate}
\item Condition~\ref{assup:converge} holds.

\item For every $P\in\mathcal P_M$, $H(P)$ is nonsingular and
$\Phi(P)\in\mathcal P_M$. Moreover, for all
$P,\hat P\in\mathcal P_M$,
\begin{equation}
\|\Phi(\hat P)-\Phi(P)\|_{\max,P}
\le\rho\|\hat P-P\|_{\max,P}.
\label{eq:pc-general-contraction}
\end{equation}

\item There is a unique fixed point $P^*\in\mathcal P_M$.
Let $K^*:=\mathcal K(P^*)$.
For the Riccati iteration starting from $P^{[0]}=\mathbf0$,
we have
\[
\|P^{[k]}-P^*\|_{\max,P}\le M\rho^k,\qquad
\|K^{[k]}-K^*\|_{\max,K}\le\frac{5b\rho M}{4r}\rho^k
\qquad(k\ge0).
\] 
The limiting feedback satisfies 
$\left\|A+\sum_{i\in\mathcal N}B^iK^{i*}\right\|_2\le\rho<1.$ 
If $s=0$, then $P^{[k]}=\mathbf0$ and $K^{[k]}=\mathbf0$
for every $k\ge0$.
\end{enumerate}
\end{theorem}

The proof of Theorem~\ref{theo:Sufficient parameter conditions}
is given in
Appendix~\ref{app:Proof of Theorem Sufficient parameter conditions}.
The inequalities in~\eqref{eq:pc-general-conditions}
provide a direct way to verify Condition~\ref{assup:converge}
and yield explicit error bounds whose coefficients and geometric
factor depend only on the model parameters.
These parameter restrictions are conservative:
Condition~\ref{assup:converge} may hold under 
weaker parameter restrictions. 

To illustrate these bounds, we give a conservative parameter range
with explicit numerical convergence estimates.
The actual range of parameters satisfying
Condition~\ref{assup:converge} is larger. 
The stated range allows arbitrary compatible matrix dimensions.

\begin{proposition}[An explicit parameter range]
\label{prop:pc-balanced}
Suppose that, for all $i\in\mathcal N$
and $j\ne i$, 
\begin{equation}
\begin{aligned}
\|A\|_2&\le0.6,
&\|C\|_2&\le0.5,\\
\sum_{i\in\mathcal N}\|B^i\|_2&\le0.06,
&\sum_{i\in\mathcal N}\|D^i\|_2&\le0.08,\\
\mathbf0\preceq Q^i&\preceq I_p,
&\frac14I_{m_i}\preceq R^{ii}&\preceq\frac13I_{m_i},\\
\mathbf0\preceq R^{ij}&\preceq\frac{0.04}{N-1}I_{m_j},
&0<\delta^i&\le1.
\end{aligned}
\label{eq:pc-balanced-conditions}
\end{equation}
Then Condition~\ref{assup:converge} holds.
Moreover, for every integer $k\ge0$,
\[
\|P^{[k]}-P^*\|_{\max,P}\le0.8(0.64)^k,\qquad
\|K^{[k]}-K^*\|_{\max,K}\le0.15(0.64)^k,
\]
and the limiting feedback satisfies 
$\left\|A+\sum_{i\in\mathcal N}B^iK^{i*}\right\|_2\le0.64<1.$ 
\end{proposition}

\begin{proof}
The bounds on the cost matrices give
\[
q\le1,\qquad r\ge\frac14,\qquad
\overline R\le\frac13+0.04=\frac{28}{75},
\]
and hence $\kappa<5/2$.
Together with the bounds on $A,B^i,C,D^i$, we obtain
\[
a\le0.6<1,\qquad
a+5\kappa\|C\|_2b\le0.975<1,\qquad
5\kappa d\le1.
\]
Thus Theorem~\ref{theo:Sufficient parameter conditions} applies,
and Condition~\ref{assup:converge} holds.

For the numerical convergence bounds, the definitions of
$s$, $M$, and $\rho$ give
\[
s\le\frac14,\qquad
M\le\frac{25}{32}<0.8,\qquad
\rho\le0.64.
\]
In particular, $\rho M\le1/2$, so
\[
\frac{5b\rho M}{4r}
\le\frac{5b}{8r}
\le\frac{5b}{2}
\le0.15.
\]
Substituting these estimates into item~3 of
Theorem~\ref{theo:Sufficient parameter conditions}
proves both convergence bounds and the closed-loop bound.
\end{proof}

\subsection{Finite-Horizon Strategies and Cost Approximation}
\label{subsec:finite-horizon-strategies}

We now study finite-horizon strategies for the infinite-horizon
game, allowing players to use different prediction horizons.
Our aim is to quantify how the resulting costs approach those
of the limiting FNE as the prediction horizons increase.
We begin by relating the Riccati iterates to the finite-horizon
FNE matrices.

Under Condition~\ref{assup:converge}(i), by Lemma~\ref{lm:Uniqueness_FNE}, let $K_t^{i*}(T)$ and
$P_t^{i*}(T)$ denote the feedback and cost matrices of the unique
FNE of the $T$-stage game~\eqref{eq:isosystem},~\eqref{eq:costfun},
with zero reference and zero terminal cost.
The corresponding strategies are
\[
u_t^{i*}=K_t^{i*}(T)x_t,\qquad
i\in\mathcal N,\quad 1\le t\le T,
\]
and $P_{T+1}^{i*}(T)=\mathbf0$ for every $i\in\mathcal N$.

Recall that $P^{[T-t]}$ is obtained by applying $\Phi$
$T-t$ times, starting from $P^{[0]}=\mathbf0$. Since the finite-horizon recursion starts from zero terminal cost 
and applies the same maps $\mathcal K$ and $\Phi$, we have
\[
P_t^{i*}(T)=\Phi^i(P^{[T-t]}),\qquad
K_t^{i*}(T)=\mathcal K^i(P^{[T-t]}),\qquad 1\le t\le T.
\] 
These identities allow us to express the iteration bounds
directly in terms of the horizon length.
For example, under the parameter conditions of
Proposition~\ref{prop:pc-balanced}, we obtain
\[
\|P_1^{i*}(T)-P^{i*}\|_2\le0.8(0.64)^T,\qquad
\|K_1^{i*}(T)-K^{i*}\|_2\le0.15(0.64)^{T-1}
\]
for every $i\in\mathcal N$ and $T\ge1$.
Thus, the first-stage cost and feedback matrices admit explicit
geometric error bounds in the horizon length.
Table~\ref{tab:Notation} summarizes the notation for the
finite-horizon matrices, the iterates, and their limits.

\begin{center}
{\small
    \refstepcounter{table}
    \label{tab:Notation}
    \centerline{\small {\bf Table \thetable}~~Notation}\vskip 1mm

    \begin{tabular}{@{}l@{\hspace{8mm}}l@{}}
        \toprule
        Notation & Description \\[0.8ex]
        \midrule
        $K_t^{i*}(T)$
        & Feedback matrix of the $T$-stage FNE, $1\le t\le T$ \\[0.8ex]
        $P_t^{i*}(T)$
        & Cost matrix of the $T$-stage FNE, $1\le t\le T+1$ \\[0.8ex]
        $K^{[k]}$
        & Stacked feedback matrix at the coupled Riccati iteration $k\ge0$ \\[0.8ex]
        $P^{[k]}$
        & Stacked cost matrix at the coupled Riccati iteration $k\ge0$ \\[0.8ex]
        $K^{i*}$
        & The $i$th row block of $K^*=\lim_{k\to\infty}K^{[k]}$ \\[0.8ex]
        $P^{i*}$
        & The $i$th row block of $P^*=\lim_{k\to\infty}P^{[k]}$ \\[0.8ex]
        \bottomrule
    \end{tabular}
}
\end{center}

The following lemma connects the finite-horizon coupled Riccati
difference equations~\eqref{eq:K} and~\eqref{eq:P} to the
infinite-horizon game. 

\begin{lemma}[{\citep[Remark~6.5]{1998basarNoncooperativeGame}
and \citep[Theorem~3.2]{siam2024}}]
\label{lm:convergence}
Consider the infinite-horizon game~\eqref{eq:isosystem},
\eqref{eq:costfun_infinite}.
Suppose that Condition~\ref{assup:converge} holds.
Then, for every fixed $t\in\mathbb N_+$ and every
$i\in\mathcal N$,
\[
\lim_{T\to\infty}K_t^{i*}(T)=K^{i*},\qquad
\lim_{T\to\infty}P_t^{i*}(T)=P^{i*}.
\]
Moreover, the stationary strategy profile
\[
u_t^i=K^{i*}x_t,\qquad
i\in\mathcal N,\quad t\in\mathbb N_+,
\]
constitutes an FNE of the infinite-horizon game, with
player $i$'s cost given by
\[
J^i(x_1)=\frac12x_1^\top P^{i*}x_1.
\]
\end{lemma}

The proof of Lemma~\ref{lm:convergence} is given in
Appendix~\ref{app:Proof of Lemma}. Lemma~\ref{lm:convergence} states that, as the horizon tends to
infinity, the feedback and cost matrices generated from zero
terminal cost converge to $K^*$ and $P^*$, which determine
a stationary FNE and its associated costs. 
Although the infinite-horizon game may admit multiple FNEs,
our approximation concerns the equilibrium selected by the
Riccati iteration starting from $P^{[0]}=\mathbf 0$.


Now we formally present the definition of the finite-horizon strategy. 

\begin{definition}[Finite-horizon strategy]
Consider the infinite-horizon game~\eqref{eq:isosystem},
\eqref{eq:costfun_infinite}, and suppose that
Condition~\ref{assup:converge} holds.
Player $i\in\mathcal N$ is said to adopt a finite-horizon strategy if:
\begin{enumerate}
\item Before the game begins, player $i$ chooses a prediction
horizon $T^i\in\mathbb N_+$. This horizon remains fixed throughout
the game and is not observed by the other players.

\item At each stage $t\in\mathbb N_+$, player $i$ individually
computes the unique FNE of an auxiliary $T^i$-stage game
starting from the current state $x_t$, with the same players,
dynamics~\eqref{eq:isosystem}, and cost parameters in~\eqref{eq:costfun_infinite}, and with zero reference and
zero terminal cost. Player $i$ then implements only the
corresponding first-stage equilibrium action:
\[
u_t^i(x_t)=K_1^{i*}(T^i)x_t.
\]
\end{enumerate}
\end{definition}

The feedback matrices $K_1^{i*}(T^i)$ depend only on the
model parameters and $T^i$, not on the current state.
More precisely, the auxiliary game is the $T^i$-stage
game of Section~2 with zero reference and zero terminal
cost; the only difference is that the initial state is
$ {x}_t$ rather than $ {x}_1$. 
Its equilibrium feedback matrices are therefore the same
as those defined above. Since the model parameters and
prediction horizons remain fixed, each player can compute
$K_1^{i*}(T^i)$ once before play. At each stage,
implementation then requires only evaluating
$  u_t^i=K_1^{i*}(T^i)  x_t$
at the observed state.

We now study the costs incurred when all players adopt these finite-horizon 
strategies, allowing them to use different prediction horizons.
The following theorem establishes stability and cost convergence
under the finite-horizon strategies. 
Under the explicit parameter conditions of
Theorem~\ref{theo:Sufficient parameter conditions}, stability holds for every choice
of positive individual prediction horizons, and the cost gap
admits an explicit bound that decreases geometrically with
the shortest horizon.
This quantitative connection between finite-horizon strategies
and infinite-horizon equilibrium costs is a main contribution
of the paper.

\begin{theorem}[Finite-horizon strategies and cost convergence]
\label{theo:finite-horizon strategy}  
Consider the game~\eqref{eq:isosystem},
\eqref{eq:costfun_infinite} under
Condition~\ref{assup:converge}.
Suppose that all players adopt finite-horizon strategies
with fixed individual prediction horizons $T^i\in\mathbb N_+$:
\[
u_t^i=K_1^{i*}(T^i)x_t,
\qquad i\in\mathcal N,\quad t\in\mathbb N_+.
\]
Define
\[
T_h:=\min_{i\in\mathcal N}T^i,
\qquad
\epsilon:=\max_{i\in\mathcal N}
\bigl\|K_1^{i*}(T^i)-K^{i*}\bigr\|_2.
\]
Let $\tilde J^i(x_1;T^1,\ldots,T^N)$ denote the resulting cost,
and let $J^i(x_1)=\frac12x_1^\top P^{i*}x_1$ denote the cost
under the limiting FNE strategies
$u_t^i=K^{i*}x_t$, $i\in\mathcal N$.
Then the following statements hold.

\begin{enumerate}
\item
Stability and cost convergence.
There exist $T_0\in\mathbb N_+$ and $c>0$ such that,
whenever $T_h\ge T_0$, the closed-loop system under the
finite-horizon strategies is stable, and 
for each $i\in\mathcal N$, let
$\tilde P^i=\tilde P^i(T^1,\ldots,T^N)$ be the unique
symmetric matrix satisfying
\[
\tilde J^i(x_1;T^1,\ldots,T^N)
=\frac12x_1^\top\tilde P^i x_1,
\qquad x_1\in\mathbb R^n.
\]
The cost matrices satisfy
\[
\max_{i\in\mathcal N}
\bigl\|\tilde P^i-P^{i*}\bigr\|_2
\le c\epsilon.
\]
Consequently, for every $i\in\mathcal N$ and
$x_1\in\mathbb R^n$,
\[
\bigl|
\tilde J^i(x_1;T^1,\ldots,T^N)-J^i(x_1)
\bigr|
\le \frac c2\|x_1\|_2^2\epsilon. 
\]
Since $\epsilon \to 0$ as $T_h\to\infty$, it follows that
\[
\lim_{T_h\to\infty}
\tilde J^i(x_1;T^1,\ldots,T^N)
=J^i(x_1).
\]

\item 
Explicit convergence bounds.
Under the additional parameter
conditions~\eqref{eq:pc-general-conditions},
\[
\bigg\|
A+\sum_{j\in\mathcal N}B^jK_1^{j*}(T^j)
\bigg\|_2
\le \rho<1.
\]
Moreover, for every $i\in\mathcal N$,
\begin{align*}
\bigl\|K_1^{i*}(T^i)-K^{i*}\bigr\|_2
\le \frac{5bM}{4r}\rho^{T^i},\quad
\bigl\|P_1^{i*}(T^i)-P^{i*}\bigr\|_2
\le M\rho^{T^i}.
\end{align*}
The resulting cost matrices satisfy
\[
\max_{i\in\mathcal N}
\bigl\|\tilde P^i-P^{i*}\bigr\|_2
\le M\rho^{T_h}.
\]
Consequently, for every $i\in\mathcal N$ and
$x_1\in\mathbb R^n$,
\[
\bigl|
\tilde J^i(x_1;T^1,\ldots,T^N)-J^i(x_1)
\bigr|
\le \frac M2\|x_1\|_2^2\rho^{T_h}.
\]
\end{enumerate}
\end{theorem}

The proof of Theorem~\ref{theo:finite-horizon strategy}
is given in Appendix~\ref{app:Proof of Theorem}.
Under Condition~\ref{assup:converge}, finite-horizon
strategies with possibly different prediction horizons
stabilize the system when $T_h=\min_{i\in\N}T^i$
is sufficiently large, and the resulting cost matrices
differ from those of the limiting FNE by at most $c\epsilon$.
By Lemma~\ref{lm:convergence}, $\epsilon\to0$ as
$T_h\to\infty$, so each player's induced cost converges
to the corresponding cost under the limiting FNE.
When the infinite-horizon game admits multiple FNEs,
this construction also provides an equilibrium selection
procedure: the finite-horizon strategies approximate
the particular FNE selected by the Riccati iteration
starting from zero terminal cost.

Under the additional parameter conditions of
Theorem~\ref{theo:Sufficient parameter conditions},
stability holds for every choice of positive prediction
horizons, and
\[
\left|
\tilde J^i(x_1;T^1,\ldots,T^N)-J^i(x_1)
\right|
\le
\frac{M}{2}\|x_1\|_2^2\rho^{T_h},
\qquad i\in\N.
\]
Since $M \ge 0$ and $0<\rho<1$  depend only on the model
parameters, each player can compute them and $\|x_1\|_2$
\emph{before the game begins}, without solving for the
limiting FNE.
For any prescribed cost tolerance $\eta>0$, the smallest
positive prediction horizon certified by this bound is
\[
T(\eta)
:=
\max\left\{
1,\,
\left\lceil
\frac{\log\!\bigl(M\|x_1\|_2^2/(2\eta)\bigr)}
     {\log(1/\rho)}
\right\rceil
\right\},
\]
with $T(\eta):=1$ when $M\|x_1\|_2^2=0$.
Thus, each player can determine a conservative horizon
requirement in advance: if $T^i\ge T(\eta)$ for every
$i\in\N$, then every player's cost differs from its
limiting FNE cost by at most $\eta$.

\section{Numerical Simulations}
\label{sec:simulation}

We present two four-player examples with three states, two outputs,
and one control input per player. The first example examines the effect of
individual prediction horizons. The second example satisfies the parameter
conditions of Theorem~\ref{theo:Sufficient parameter conditions}, admits infinitely many
stabilizing FNEs, and examines the effect of player-specific discount
factors. Both examples use nonzero constant reference trajectories.
For their FNEs, we consider stationary affine feedbacks and allow
unilateral deviations that keep the closed-loop dynamics Schur stable.

The system dynamics and player $i$'s cost are
\begin{equation}
\begin{aligned}
x_{t+1}&=Ax_t+Bu_t,\\
y_t&=Cx_t+Du_t,
\end{aligned}
\label{eq:sim-system}
\end{equation}
\begin{equation}
J^i(x_1,u)
=\frac12\sum_{t=1}^{\infty}(\delta^i)^{t-1}
\left[
(y_t-l^i)^\top Q^i(y_t-l^i)
+\sum_{j=1}^{4}R^{ij}(u_t^j)^2
\right],
\label{eq:sim-cost}
\end{equation}
where $x_t\in\mathbb R^3$, $y_t,l^i\in\mathbb R^2$, and
\[
u_t=\operatorname{col}(u_t^1,u_t^2,u_t^3,u_t^4)
\in\mathbb R^4,
\qquad u_t^i\in\mathbb R.
\]
Each player uses the finite-horizon strategy
\[
u_t^i=K_1^{i*}(T^i)x_t+L_1^{i*}(T^i),
\]
where $K_1^{i*}(T^i)\in\mathbb R^{1\times3}$ and
$L_1^{i*}(T^i)\in\mathbb R$ are computed by backward
recursion for the auxiliary $T^i$-stage game with
zero terminal cost. We compute costs from $5000$-stage trajectories and bound the
remaining discounted tails.  Since the references are nonzero, these cost experiments
extend beyond the zero-reference setting of the theoretical analysis.

\subsection{Individual Prediction Horizons}

In the first example, we use
\[
A=\begin{bmatrix}
0.3&0&-0.2\\
0.2&0.4&0.1\\
-0.2&0.3&0.5
\end{bmatrix},\qquad
B=\begin{bmatrix}
0.2&0.5&-0.3&-0.5\\
0.9&0.4&0.6&0.5\\
-0.3&0.6&0.4&-1.2
\end{bmatrix},
\]
\[
C=\begin{bmatrix}
0.3&-0.2&0.5\\
0.4&0.1&-0.7
\end{bmatrix},\qquad
D=\begin{bmatrix}
0.6&-0.1&-0.3&0.8\\
0.2&0.3&0.4&-0.4
\end{bmatrix}.
\]
The cost parameters are
\[
Q^1=4I_2,\qquad Q^2=2I_2,\qquad Q^3=2.5I_2,
\qquad Q^4=\operatorname{diag}(1.3,1),
\]
\[
(R^{ij})_{i,j=1}^{4}=\begin{bmatrix}
0.70&0.10&0.10&0.88\\
0.20&0.50&0.10&0.03\\
0.10&0.10&0.90&0.10\\
0.09&1.14&0.10&0.94
\end{bmatrix},\qquad
(\delta^1,\delta^2,\delta^3,\delta^4)=(0.9,0.6,0.95,0.98).
\]
The references and initial state are
\[
\begin{bmatrix}  l^1&  l^2&  l^3&  l^4\end{bmatrix}
=\begin{bmatrix}1&-1&-0.5&1\\1&-1&0.5&-1\end{bmatrix},\qquad
  x_1=
\begin{bmatrix}-0.353\\-1.926\\-2.595\end{bmatrix}.
\]
This example does not satisfy the sufficient parameter conditions of
Theorem~\ref{theo:Sufficient parameter conditions}. We use it to explore the behavior of the
finite-horizon strategies outside that parameter range.

We use $5$ and $200$ as the short and long horizons, respectively.
The horizons of players 1--3 are fixed at
\[
(T^1,T^2,T^3)\in
\{(200,200,200),(5,200,200),(5,5,200),(5,5,5)\},
\]
and $T^4$ varies from $2$ to $400$. Let $s\in\{0,1,2,3\}$ denote the
number of short-horizon players among players 1--3, assigned in the
order 1, 2, 3.

We continue the coupled recursion to its limit and check the Bellman
and first-order equations, the positive own-control Hessians, and
closed-loop stability. The limiting affine FNE has closed-loop
spectral radius $0.721980$, and its costs are
\[
\begin{aligned}
J^1(  x_1)&=261.081,&
J^2(  x_1)&=37.149,\\
J^3(  x_1)&=304.237,&
J^4(  x_1)&=1016.023.
\end{aligned}
\]
The profile $(T^1,T^2,T^3,T^4)=(5,5,5,2)$ has closed-loop spectral
radius $1.025950$ and is omitted on stability grounds. All other
$1595$ tested profiles are stable.

\begin{figure}
\centering
\includegraphics[width=0.84\textwidth]{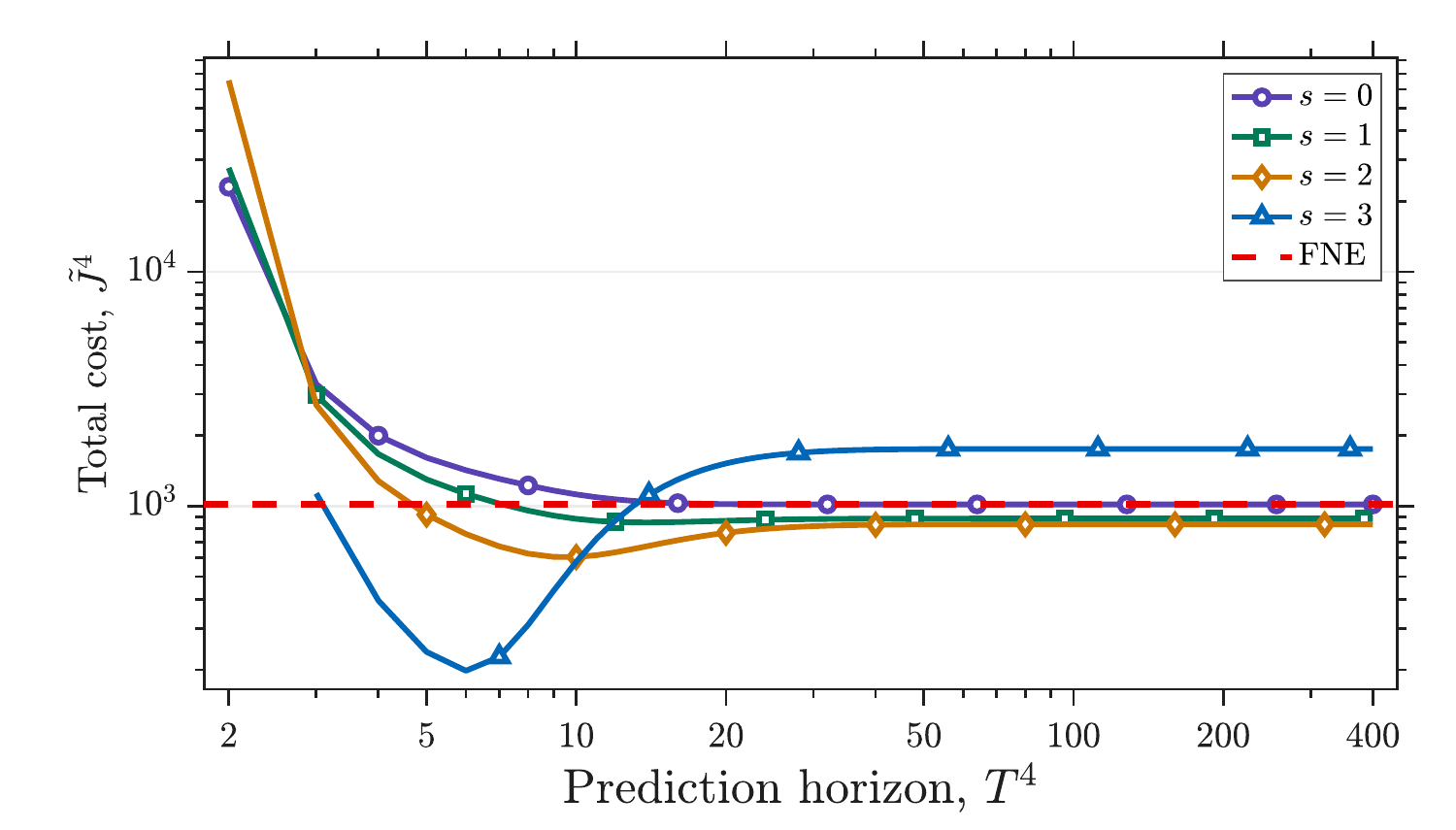}
\caption{Player 4's infinite-horizon cost under the four specified
horizon profiles. The red dashed line gives its cost under the limiting
affine FNE. Both axes use logarithmic scales. The unstable profile $(5,5,5,2)$
is omitted.}
\label{fig:sim-player4}
\end{figure}

Figure~\ref{fig:sim-player4} illustrates two distinct effects.
First, when the other three players use the long horizon,
player 4's cost is close to the FNE baseline (the red dashed line) when its own
horizon is also large. If some opponents use short horizons,
however, player 4's long-horizon cost may be either
higher or lower. 
Second, increasing player 4's horizon may not reduce its
cost, so the longest tested horizon may not be the best
choice. For $(T^1,T^2,T^3)=(5,5,200)$, the lowest cost over
the tested horizons is $606.039$ at $T^4=10$, whereas the
cost at $T^4=400$ is $835.193$. For
$(T^1,T^2,T^3)=(5,5,5)$, the lowest tested cost is $198.170$
at $T^4=6$, compared with $1751.839$ at $T^4=400$.
Thus, with short-horizon opponents, a longer individual
horizon neither guarantees a lower cost than the FNE
baseline nor necessarily improves on a shorter horizon.

\begin{figure}
\centering
\includegraphics[width=\textwidth]{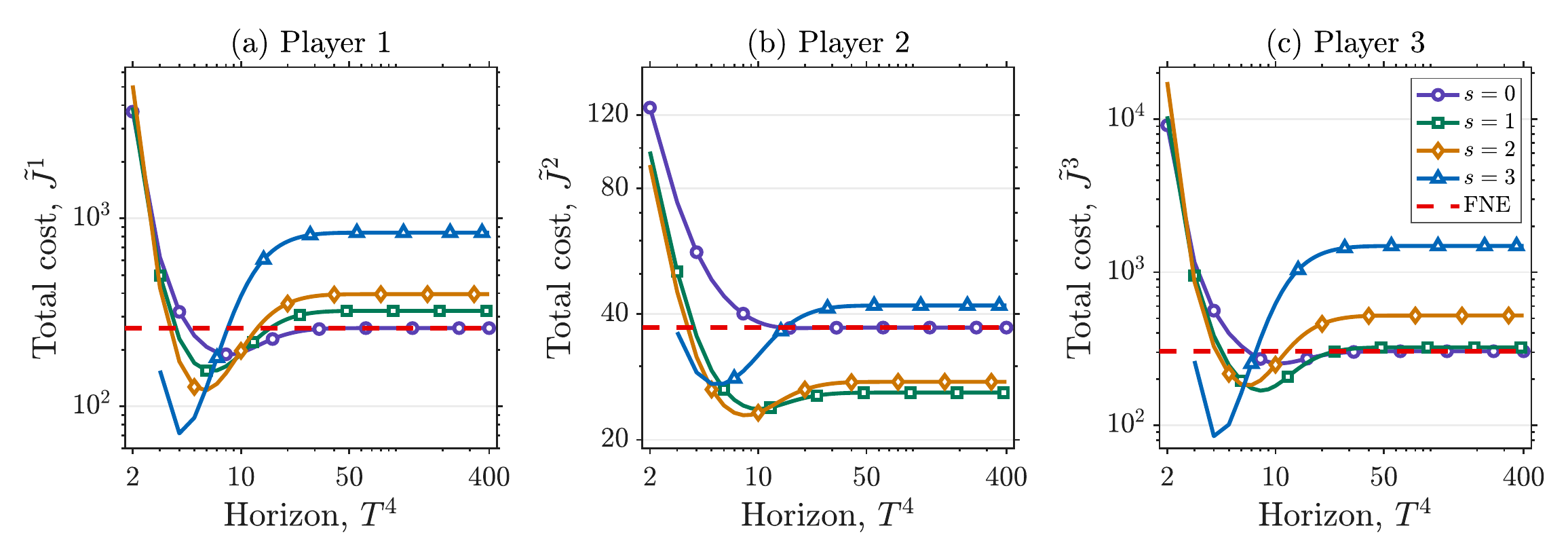}
\caption{Infinite-horizon costs of players 1--3 under the same horizon
profiles as Figure~\ref{fig:sim-player4}. Each red dashed line gives that
player's limiting affine FNE cost. Both axes use logarithmic scales.
Only stable profiles are shown.}
\label{fig:sim-others}
\end{figure}

Figure~\ref{fig:sim-others} shows that changing $T^4$ also
affects the other players. For $(T^1,T^2,T^3)=(5,5,5)$,
all four players have higher costs at $T^4=400$ than at
$T^4=6$. Thus, when players 1--3 use these short horizons,
player 4 is better off also using a short horizon in this
comparison, and the other three players benefit as well.
The number of short-horizon opponents alone does not
determine whether a player's cost rises or falls.

These observations are consistent with our theoretical results,
which concern approximation of an infinite-horizon FNE as all
prediction horizons increase. In this example, the horizon profile
$(T^1,T^2,T^3,T^4)=(5,5,5,6)$ yields strictly lower costs for all
four players than the limiting FNE, and thus provides a strict
Pareto improvement over it. Our theoretical results, however,
do not provide general conditions for the existence of such
improvements or characterize Pareto-optimal combinations of
players' prediction horizons.

\subsection{Multiple FNEs and Player-Specific Patience}

The second example uses
\[
A=\begin{bmatrix}
0.95&0&0.01\\
0&0.95&0.01\\
0&0&0.75
\end{bmatrix},\qquad
B=0.01\begin{bmatrix}
1&0&0.8&-0.6\\
0&1&0.6&0.8\\
0&0&0&0
\end{bmatrix},
\]
\[
C=0.01\begin{bmatrix}
1.96&0&0.5\\
0&1.96&0.5
\end{bmatrix},\qquad
D=-0.01\begin{bmatrix}
0.2&0&0.16&-0.12\\
0&0.2&0.12&0.16
\end{bmatrix}.
\]
For every player, $Q^i=I_2$, and
\[
(R^{ij})_{i,j=1}^{4}=0.1\begin{bmatrix}
1&1&5&5\\
1&1&5&5\\
5&5&1&1\\
5&5&1&1
\end{bmatrix}.
\]
The references and initial state are
\[
\begin{bmatrix}  l^1&  l^2&  l^3&  l^4\end{bmatrix}
=\begin{bmatrix}
2.68&0&4.77&-0.03\\
0&2.68&3.58&-0.04
\end{bmatrix},\qquad
  x_1=
\begin{bmatrix}10.699\\9.700\\14.810\end{bmatrix}.
\]
We first take $\delta^i=0.99$ for all players. The parameter conditions
of Theorem~\ref{theo:Sufficient parameter conditions} hold because
\[
a=0.950<1,\qquad
a+5\kappa\|C\|_2b=0.996<1,\qquad
5\kappa d=0.438<1.
\]
The theorem gives $M=0.009$ and $\rho=0.955$.
Thus, $H$ is nonsingular on $\mathcal P_M$, and the zero-terminal
quadratic Riccati iteration converges geometrically to the unique
fixed point in that set. These conclusions concern the quadratic cost and feedback
matrices, which do not depend on the reference trajectories.

Nevertheless, the game with the stated nonzero references admits
infinitely many stabilizing affine FNEs. Appendix~\ref{app:simulation-fnes}
constructs such a family of FNEs. There is no conflict since
the zero-terminal iteration just selects one limiting FNE in all infinite-horizon FNEs.

We also check the convergence of the affine terms and verify the
limiting strategy using the Bellman equations and first-order
conditions. Its closed-loop spectral radius is $0.950$. Its costs are
\[
\begin{aligned}
J^1(  x_1)&=349.050,&
J^2(  x_1)&=349.937,\\
J^3(  x_1)&=1746.767,&
J^4(  x_1)&=1.006.
\end{aligned}
\]
The tracking error term accounts for about $91.4\%$ of player 4's cost.

\begin{figure}
\centering
\includegraphics[width=\textwidth]{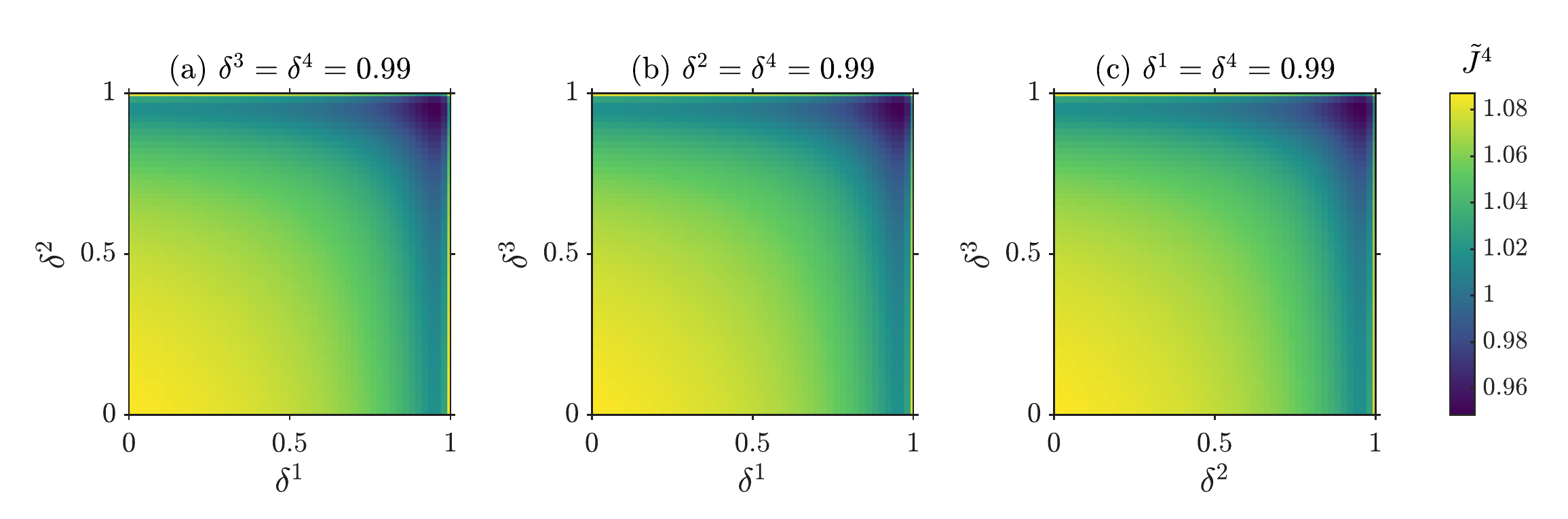}
\caption{Player 4's infinite-horizon cost as two opponents' discount
factors vary. The fixed pairs are (a) players 3 and 4, (b) players 2
and 4, and (c) players 1 and 4, each with discount factor $0.99$.
All prediction horizons equal $200$, and all panels use the same cost scale.}
\label{fig:sim-patience}
\end{figure}

To examine patience, we set $T^i=200$ for all players. In each panel,
two discount factors are fixed at $0.99$, and the other two vary
independently over $\{0,0.02,\ldots,1\}$:
\[
\begin{array}{c|c|c}
\text{Panel}&\text{Fixed discount factors}&\text{Varied discount factors}\\ \hline
\text{(a)}&\delta^3=\delta^4=0.99&(\delta^1,\delta^2)\\
\text{(b)}&\delta^2=\delta^4=0.99&(\delta^1,\delta^3)\\
\text{(c)}&\delta^1=\delta^4=0.99&(\delta^2,\delta^3)
\end{array}
\]
At each grid point, we recompute the auxiliary 200-stage FNE,
interpreting $\delta^i=0$ as the limit $\delta^i\rightarrow0$.  Figure~\ref{fig:sim-patience} shows player 4's costs under each profile of patience and  uses a common color scale. The full range in each panel is
about $0.949$ to $1.087$. Holding one's discount factor fixed,
changing the other discount factor can change the cost by about $0.069$, or $6.9\%$
of player 4's baseline FNE cost. The cost initially decreases with patience
but rises again near one. Thus, greater patience can help or harm
another player, even when all players use a long horizon. This experiment illustrates a
possible effect of patience; it does not give a general monotonicity
result for LQ games.

\section{Conclusions}
\label{sec:conclusions}
In this paper, we study discrete-time infinite-horizon LQ dynamic games with i/o/s dynamics. We introduce a finite-horizon strategy, under which each player solves an auxiliary finite-horizon game with an individual prediction horizon and implements only the first-stage equilibrium control. Our theoretical analysis focuses on the zero-reference case. We give explicit conditions on the model parameters under which the coupled Riccati iteration is well defined and converges geometrically to a stabilizing solution. Under these conditions, the finite-horizon strategies stabilize the system for any positive individual prediction horizons, and each player's total cost converges to that under the limiting infinite-horizon FNE as all prediction horizons tend to infinity. Moreover, we derive an explicit upper bound on this cost gap that decreases geometrically with the shortest prediction horizon. This bound tells us how long the prediction horizons need to be for a given accuracy, without directly solving the coupled algebraic Riccati equations. Although the theoretical results are established for the
zero-reference case, we provide two non-scalar numerical
examples with nonzero output references to illustrate
the proposed strategy.

The parameter conditions obtained in this paper are sufficient but conservative. Future work will seek weaker parameter  conditions and extend the convergence analysis to nonzero reference trajectories.

\section*{Declaration of AI Use}
The authors used
ChatGPT for language polishing and to assist
with a limited portion of the mathematical reasoning and proofs.
The authors have verified the content and take full responsibility
for the accuracy and integrity of the manuscript.

\bibliographystyle{unsrt}
\bibliography{main}

\appendix

\section{Reformulation of the i/o/s Model as an i/s LQ Game}
\label{app:ios_equivalence}

Before we show how the finite-horizon i/o/s game considered in this paper can be rewritten as an i/s LQ dynamic game with time-invariant dynamics and time-varying cost matrices, we first  clarify the inclusion relations between our model and i/s LQ games. 

Let $\mathcal S_0$ denote time-invariant i/s LQ games without cross terms.
Let $\mathcal S_1$ denote time-invariant i/s LQ games allowing cross terms.
Let $\mathcal S_2$ denote i/s LQ games with time-invariant dynamics, allowing time-varying costs, cross terms, linear terms, and constant terms.
Here, cross, linear, and constant terms all refer to the cost functions. Cross terms are products between states and controls, or between different players' controls.
Let $\mathcal O_0$ denote our i/o/s games with $l_t^i=0$ for all players and stages, and let $\mathcal O$ denote our i/o/s games allowing non-zero reference trajectories $l_t^i$. Then we have 
\[
\begin{gathered}
\mathcal S_0\subsetneq\mathcal O_0\subsetneq\mathcal S_1\subsetneq\mathcal S_2,\\
\mathcal O_0\subsetneq\mathcal O\subsetneq\mathcal S_2,\\
\mathcal O\nsubseteq\mathcal S_1,\qquad \mathcal S_1\nsubseteq\mathcal O.
\end{gathered}
\]

Setting $C=I$, $D=0$, and $l_t^i=0$ gives the games in $\mathcal S_0$. Nonzero $D$ can introduce cross terms. With zero references, the cost coefficients remain constant, so our games belong to $\mathcal S_1$. However, our model does not cover all costs in $\mathcal S_1$. For example, the scalar cost $(x-u^i)^2=x^2-2xu^i+(u^i)^2$ is allowed in $\mathcal S_1$: it is nonnegative, and the coefficient of $(u^i)^2$ is positive. Yet it equals zero at $x=u^i\ne0$. In our model, the separate control cost $(u^i)^\top R^{ii}u^i$ is strictly positive whenever $u^i\ne0$. Nonzero references can add linear and constant terms that vary with time. Thus, some games in $\mathcal O$ fall outside $\mathcal S_1$. The references do not change the quadratic coefficients, while $\mathcal S_2$ allows these coefficients to vary as well. Hence, $\mathcal O$ is a proper subclass of $\mathcal S_2$. We now give the equivalent i/s formulation.

Define $R^i=
    \text{diag}(R^{i1},\ldots,R^{iN}).$ 
Substituting \(y_t=Cx_t+Du_t\) into the stage cost gives
\begin{align*}
    J^i
    =&
    \frac{1}{2}
    \sum_{t=1}^{T}
    (\delta^i)^{t-1}
    \Bigg[
    x_t^\top C^\top Q^i C x_t
    +2x_t^\top C^\top Q^i D u_t + u_t^\top (D^\top Q^i D+R^i)u_t    -2(l_t^i)^\top Q^i C x_t
    \\ &-2(l_t^i)^\top Q^i D u_t
    +(l_t^i)^\top Q^i l_t^i
    \Bigg].
    \label{eq:ios_expanded_cost}
\end{align*} 
Now define the augmented state
$z_t =
    \begin{bmatrix}
        x_t\\
        1
    \end{bmatrix}.$ 
Then \(z_t\) satisfies the standard i/s dynamics 
    $z_{t+1}
    =
    \bar A z_t+\bar B u_t, $ where $
    \bar A
    :=
    \begin{bmatrix}
        A & 0\\
        0 & 1
    \end{bmatrix}, 
    \bar B
    :=
    \begin{bmatrix}
        B\\
        0
    \end{bmatrix}.$ 
For each player \(i\), define
\begin{equation*}
    \bar Q_t^i
    :=
    (\delta^i)^{t-1}
    \begin{bmatrix}
        C^\top Q^i C        & -C^\top Q^i l_t^i\\
        -(l_t^i)^\top Q^i C & (l_t^i)^\top Q^i l_t^i
    \end{bmatrix},
\end{equation*} 
\begin{equation*}
    \bar S_t^i
    :=
    (\delta^i)^{t-1}
    \begin{bmatrix}
        C^\top Q^i D\\
        -(l_t^i)^\top Q^i D
    \end{bmatrix},
    \quad
    \bar R_t^i
    :=
    (\delta^i)^{t-1}
    \left(D^\top Q^i D+R^i\right).
\end{equation*}
With these definitions, the expanded cost in
\eqref{eq:costfun} is exactly
\begin{equation}
    J^i =
    \frac{1}{2}
    \sum_{t=1}^{T}
    \bigg[
        z_t^\top \bar Q_t^i z_t
        +2z_t^\top \bar S_t^i u_t
        +u_t^\top \bar R_t^i u_t
    \bigg].
    \label{eq:standard_is_cost}
\end{equation}
Therefore, the original discounted output-tracking game in this paper is equivalent to the
standard finite-horizon LQ dynamic game
\begin{equation*}
    z_{t+1}=\bar A z_t+\bar B u_t,
\end{equation*}
with player-\(i\) cost functional given by \eqref{eq:standard_is_cost}.

\section{Proof of Lemma~\ref{lm:Uniqueness_FNE}}
\label{app:lm:Uniqueness_FN}

\begin{proof}
We proceed by backward induction.
The terminal coefficients are uniquely specified by
$P^i_{T+1}=\boldsymbol 0$, $S^i_{T+1}=\boldsymbol 0$, and $w^i_{T+1}=0$
for all $i\in\mathcal N$.

Suppose that $P^i_{t+1}$, $S^i_{t+1}$, and $w^i_{t+1}$
have been uniquely determined for all $i\in\mathcal N$.
Since $H(P_{t+1})$ is nonsingular by assumption,
the feedback matrices $K_t$ and affine terms $L_t$
are uniquely determined by
\[
K_t=\mathcal K(P_{t+1}),\qquad
L_t=H(P_{t+1})^{-1}\tilde g(S_{t+1}).
\] 
Substituting $K_t$ and $L_t$ into
\eqref{eq:P}--\eqref{eq:w} uniquely determines
$P^i_t$, $S^i_t$, and $w^i_t$ for every $i\in\mathcal N$.

Applying this argument successively for $t=T,T-1,\ldots,1$
uniquely determines
$\{P^i_t,S^i_t,w^i_t\}_{t=1}^{T+1}$ for all
$i\in\mathcal N$, together with
$\{K_t,L_t\}_{t=1}^{T}$.
These coefficients form the unique solution to
\eqref{eq:K}--\eqref{eq:w} with the prescribed terminal
conditions. The existence and uniqueness of the FNE
then follow from Lemma~\ref{lm:suffi nece}.
\end{proof}

\section{Proof of Theorem~\ref{theo:Sufficient parameter conditions}}
\label{app:Proof of Theorem Sufficient parameter conditions}

\begin{proof} 
We establish the properties in item~2 and then derive
items~3 and~1. All estimates also cover $s=0$.

\smallskip
\noindent\textbf{1. Invertibility of $H(P)$.} Recall the parameter definitions
\[
\begin{aligned}
a&=\|A\|_2,
& b&=\sum_{j\in\mathcal N}\|B^j\|_2,
& d&=\sum_{j\in\mathcal N}\|D^j\|_2,\\
q&=\max_{i\in\mathcal N}\|Q^i\|_2,
& r&=\min_{i\in\mathcal N}\lambda_{\min}(R^{ii}),
& \overline R&=\max_{i\in\mathcal N}
                 \sum_{j\in\mathcal N}\|R^{ij}\|_2,
\end{aligned}
\]
and
\[
\kappa=\frac{\sqrt{q\overline R}}{r},
\qquad
s=q\|C\|_2^2,
\qquad
M=\frac{2s}{1-a^2}.
\]
We have $0<r\le\overline R$, and the parameter conditions imply
\[
0\le a<1,\qquad
\kappa\|C\|_2b\le\frac{1-a}{5},
\qquad
\kappa d\le\frac15.
\] 
The parameter conditions give
\[
\frac{qd^2+Mb^2}{r}
=\frac r{\overline R}
\left[(\kappa d)^2+\frac{2(\kappa\|C\|_2b)^2}{1-a^2}\right]
\le\frac1{25}\left(1+\frac{2(1-a)}{1+a}\right) 
<\frac15.
\]
Fix $P\in\mathcal P_M$. Let
$Z=\operatorname{col}(Z^1,\ldots,Z^N)$, where
$Z^j\in\mathbb R^{m_j\times n}$, and set $E=H(P)Z$.
Choose $i\in\mathcal N$ such that
$\|Z^i\|_2=\|Z\|_{\max,K}$. 
Recall that the $(i,j)$ block of $H(P)$ is
$(D^i)^\top Q^iD^j+\delta^i(B^i)^\top P^iB^j$,
with the additional term $R^{ii}$ when $j=i$.
Thus, the $i$th block of $E=H(P)Z$ is 
\[
E^i
=R^{ii}Z^i
+\sum_{j\in\mathcal N}
\bigl[(D^i)^\top Q^iD^j
+\delta^i(B^i)^\top P^iB^j\bigr]Z^j.
\]
Thus, 
\[
R^{ii}Z^i
=E^i-(D^i)^\top Q^i\sum_{j\in\mathcal N}D^jZ^j
-\delta^i(B^i)^\top P^i\sum_{j\in\mathcal N}B^jZ^j.
\]
For a real symmetric matrix, the matrix $2$-norm equals
the largest absolute value of its eigenvalues.
Since $R^{ii}\succeq rI_{m_i}$ with $r>0$,
both $R^{ii}$ and its inverse are symmetric positive definite.
Consequently,
\[
\|(R^{ii})^{-1}\|_2
=\lambda_{\max}\bigl((R^{ii})^{-1}\bigr)
=\frac{1}{\lambda_{\min}(R^{ii})}
\le\frac1r.
\]
Using $\|Z^i\|_2=\|Z\|_{\max,K}$ and the block equation, we obtain
\begin{align*}
r\|Z\|_{\max,K}
&=r\|Z^i\|_2
=r\|(R^{ii})^{-1}(R^{ii}Z^i)\|_2\\
&\le r\|(R^{ii})^{-1}\|_2\|R^{ii}Z^i\|_2
\le\|R^{ii}Z^i\|_2\\
&\le\|E^i\|_2
+\|D^i\|_2\|Q^i\|_2\sum_j\|D^j\|_2\|Z^j\|_2
+\delta^i\|B^i\|_2\|P^i\|_2\sum_j\|B^j\|_2\|Z^j\|_2\\
&\le\|E\|_{\max,K}
+\bigl(q\|D^i\|_2d+M\|B^i\|_2b\bigr)\|Z\|_{\max,K}\\
&\le\|E\|_{\max,K}+(qd^2+Mb^2)\|Z\|_{\max,K}.
\end{align*}
Here we used $\|Q^i\|_2\le q$, $\delta^i\|P^i\|_2\le M$,
$\|Z^j\|_2\le\|Z\|_{\max,K}$, and the definitions of $b,d$.
Rearranging and using $qd^2+Mb^2 <r/5$ yields
\[
\|E\|_{\max,K}
\ge\bigl(r-qd^2-Mb^2\bigr)\|Z\|_{\max,K}
\ge\frac{4r}{5}\|Z\|_{\max,K}.
\]
Thus every nonzero $Z$ has $H(P)Z\ne\mathbf 0$.
Since $H(P)$ is square, it is invertible, with
\begin{equation}
\|H(P)^{-1}E\|_{\max,K}\le\frac5{4r}\|E\|_{\max,K}.
\label{eq:pc-simple-inverse}
\end{equation}
This proves the invertibility in item~2.

\smallskip
\noindent\textbf{2. Feedback and Closed-Loop Bounds.}
The parameter conditions also give
\begin{align*}
q\|C\|_2d+aMb
&=r\sqrt{\frac{s}{\overline R}}
\left[\kappa d+\frac{2a\kappa\|C\|_2b}{1-a^2}\right]
\le\frac r5\sqrt{\frac{s}{\overline R}}
\left(1+\frac{2a}{1+a}\right)
\le\frac{2r}{5}\sqrt{\frac{s}{\overline R}}.
\end{align*}
For every $i\in\mathcal N$, the definition of $g(P)$ gives
\begin{align*}
\|g^i(P)\|_2
&=\|(D^i)^\top Q^iC+\delta^i(B^i)^\top P^iA\|_2\\
&\le\|D^i\|_2\|Q^i\|_2\|C\|_2
+\delta^i\|B^i\|_2\|P^i\|_2\|A\|_2\\
&\le q\|C\|_2\|D^i\|_2+aM\|B^i\|_2\\
&\le q\|C\|_2d+aMb.
\end{align*}
Here we used $\|Q^i\|_2\le q$,
$\delta^i\|P^i\|_2\le M$, $\|A\|_2=a$,
$\|D^i\|_2\le d$, and $\|B^i\|_2\le b$.
Taking the maximum over $i$ yields
\[
\|g(P)\|_{\max,K}
=\max_{i\in\mathcal N}\|g^i(P)\|_2
\le q\|C\|_2d+aMb.
\] 
Using $\|g(P)\|_{\max,K}\le q\|C\|_2d+aMb$ and
\eqref{eq:pc-simple-inverse} gives
$\|\mathcal K(P)\|_{\max,K}\le\tfrac12\sqrt{s/\overline R}$.
Using the definitions of $F,G$, the identity
$\sqrt{s/\overline R}=(r/\overline R)\kappa\|C\|_2$,
and $r/\overline R\le1$, we obtain
\begin{equation}
\begin{aligned}
&\|\mathcal K(P)\|_{\max,K}
\le\frac12\sqrt{\frac{s}{\overline R}},\\
&\|F(P)\|_2
\le a+b\|\mathcal K(P)\|_{\max,K}
\le a+\frac{r}{2\overline R}\kappa\|C\|_2b \le a+\frac{1-a}{10}=\rho,\\
&\|G(P)\|_2
\le\|C\|_2+d\|\mathcal K(P)\|_{\max,K}
\le\left(1+\frac{r}{2\overline R}\kappa d\right)\|C\|_2 \le\frac{11}{10}\|C\|_2.
\end{aligned}
\label{eq:pc-simple-bounds}
\end{equation}
Here the final inequalities follow from
$\kappa\|C\|_2b\le(1-a)/5$ and $\kappa d\le1/5$.

\smallskip
\noindent\textbf{3. Invariance of $\mathcal P_M$.}
Since $Q^i\succeq\mathbf 0$, $P^i\succeq\mathbf 0$,
$R^{ij}\succeq\mathbf 0$, and $\delta^i>0$, we have 
\[
\Phi^i(P)
=G(P)^\top Q^iG(P)
+\delta^iF(P)^\top P^iF(P)
+\sum_{j\in\mathcal N}\mathcal K^j(P)^\top R^{ij}\mathcal K^j(P)
\succeq\mathbf 0.
\] 
For the upper bound, apply \eqref{eq:pc-simple-bounds} directly
to its three quadratic terms:
\[
\|\Phi^i(P)\|_2
\le q\|G(P)\|_2^2+M\|F(P)\|_2^2
+\overline R\|\mathcal K(P)\|_{\max,K}^2
\le M\rho^2+\frac{73}{50}s.
\]
By the definitions of $M$ and $\rho$,
\[
M(\rho^2-a^2)
=\frac{s}{5(1+a)}\left(2a+\frac{1-a}{10}\right)
\le\frac s5.
\]
Consequently,
\[
\|\Phi^i(P)\|_2\le Ma^2+\frac{83}{50}s
\le Ma^2+2s=M.
\]
Together with positivity, this proves
$\Phi(\mathcal P_M)\subseteq\mathcal P_M$.

\smallskip
\noindent\textbf{4. Contraction.}
Fix $P,\hat P\in\mathcal P_M$. Let
$K=\mathcal K(P)$, $F=F(P)$, and $G=G(P)$;
use hats for evaluation at $\hat P$, and set
$\Delta P=\hat P-P$, $\Delta K=\hat K-K$.
Then $\hat F-F=B\Delta K$ and $\hat G-G=D\Delta K$.

For each $i\in\mathcal N$, the $i$th block of
$H(P)K=g(P)$ can be written as
\[
R^{ii}K^i+(D^i)^\top Q^iG
+\delta^i(B^i)^\top P^iF=0.
\]
Subtracting this equation from its counterpart at $\hat P$
and using $\hat G-G=D\Delta K$ gives
\[
R^{ii}\Delta K^i+(D^i)^\top Q^iD\Delta K
+\delta^i(B^i)^\top
(\hat P^i\hat F-P^iF)=0.
\]
Since
\[
\hat P^i\hat F-P^iF
=P^i(\hat F-F)+(\hat P^i-P^i)\hat F
=P^iB\Delta K+\Delta P^i\hat F,
\]
we obtain
\[
R^{ii}\Delta K^i
+(D^i)^\top Q^iD\Delta K
+\delta^i(B^i)^\top P^iB\Delta K
=-\delta^i(B^i)^\top\Delta P^i\hat F.
\]
The left-hand side is the $i$th block of $H(P)\Delta K$.
Stacking these equations yields
\[
H(P)\Delta K
=-\operatorname{col}_{i\in\mathcal N}
\bigl(\delta^i(B^i)^\top\Delta P^i\hat F\bigr).
\]
The inverse bound \eqref{eq:pc-simple-inverse} and
$\|\hat F\|_2\le\rho$ give
\begin{equation}
\begin{aligned}
\|\Delta K\|_{\max,K}
&\le\frac5{4r}\max_{i\in\mathcal N}
\|\delta^i(B^i)^\top\Delta P^i\hat F\|_2 \le\frac{5b\rho}{4r}\|\Delta P\|_{\max,P},\\
\sqrt{s\overline R}\,\|\Delta K\|_{\max,K}
&\le\frac{5\rho}{4}
\frac{b\sqrt{s\overline R}}{r}\|\Delta P\|_{\max,P} \le\frac{(1-a)\rho}{4}\|\Delta P\|_{\max,P}.
\end{aligned}
\label{eq:pc-simple-feedback-difference}
\end{equation}
Here we used $0<\delta^i\le1$, $\|B^i\|_2\le b$,
and $\|\Delta P^i\|_2\le\|\Delta P\|_{\max,P}$.
The last inequality follows from
\[
\frac{b\sqrt{s\overline R}}{r}
=\kappa\|C\|_2b\le\frac{1-a}{5}.
\]

Next compare the three quadratic terms in $\Phi$.
For the term involving $P^i$, we have
\[
\begin{aligned}
\|\hat F^\top\hat P^i\hat F-F^\top P^iF\|_2
&=\bigl\|
\hat F^\top\Delta P^i\hat F
+(\hat F-F)^\top P^i\hat F
+F^\top P^i(\hat F-F)
\bigr\|_2\\
&\le \rho^2\|\Delta P\|_{\max,P}
+2M\rho b\|\Delta K\|_{\max,K}.
\end{aligned}
\] 
Similarly, using $\hat G-G=D\Delta K$, we obtain
\[
\begin{aligned}
\|\hat G^\top Q^i\hat G-G^\top Q^iG\|_2
&=\|(\hat G-G)^\top Q^i\hat G
+G^\top Q^i(\hat G-G)\|_2\\
&\le q(\|\hat G\|_2+\|G\|_2)d\|\Delta K\|_{\max,K}\\
&\le\frac{11}{5}q\|C\|_2d\|\Delta K\|_{\max,K}.
\end{aligned}
\]
For the control-cost terms,
\[
\begin{aligned}
&\left\|\sum_j
\bigl[(\hat K^j)^\top R^{ij}\hat K^j
-(K^j)^\top R^{ij}K^j\bigr]\right\|_2\\
&\quad=\left\|\sum_j
\bigl[(\Delta K^j)^\top R^{ij}\hat K^j
+(K^j)^\top R^{ij}\Delta K^j\bigr]\right\|_2\\
&\quad\le\sum_j\|R^{ij}\|_2
\bigl(\|\hat K^j\|_2+\|K^j\|_2\bigr)\|\Delta K^j\|_2\\
&\quad\le\overline R\sqrt{\frac{s}{\overline R}}
\|\Delta K\|_{\max,K}
=\sqrt{s\overline R}\|\Delta K\|_{\max,K}.
\end{aligned}
\]
Combining these estimates with the bound for the term
involving $P^i$, and using $0<\delta^i\le1$, gives
\begin{align*}
\|\Phi(\hat P)-\Phi(P)\|_{\max,P}
&\le\rho^2\|\Delta P\|_{\max,P} +
2\left(\frac{11}{10}q\|C\|_2d+Mb\rho
+\frac12\sqrt{s\overline R}\right)
\|\Delta K\|_{\max,K}\\
&\le\rho^2\|\Delta P\|_{\max,P}
+3\sqrt{s\overline R}\|\Delta K\|_{\max,K}.
\end{align*}
For the last inequality, the parameter conditions
and $r\le\overline R$ imply
\[
q\|C\|_2d
=\frac r{\overline R}\kappa d\sqrt{s\overline R}
\le\frac15\sqrt{s\overline R},
\]
and
\[
Mb\rho
=\frac{2r\rho}{\overline R(1-a^2)}
\kappa\|C\|_2b\sqrt{s\overline R}
\le\frac{2\rho}{5(1+a)}\sqrt{s\overline R}
\le\frac25\sqrt{s\overline R}.
\]
Here the last inequality uses $\rho<1\le1+a$. 
Applying \eqref{eq:pc-simple-feedback-difference} now yields
\[
\|\Phi(\hat P)-\Phi(P)\|_{\max,P}
\le\rho\left(\rho+\frac34(1-a)\right)\|\Delta P\|_{\max,P}
=\rho\left(1-\frac3{20}(1-a)\right)\|\Delta P\|_{\max,P}
\le\rho\|\Delta P\|_{\max,P}.
\]
This proves contraction and completes item~2.

\smallskip
\noindent\textbf{5. Convergence, Rates, and Condition~\ref{assup:converge}.}
The set $\mathcal P_M$ is nonempty and closed in a
finite-dimensional normed space, hence complete.
Since $0<\rho<1$, invariance and contraction give a unique
fixed point $P^*\in\mathcal P_M$ and
\[
\|P^{[k]}-P^*\|_{\max,P}
\le\rho^k\|P^{[0]}-P^*\|_{\max,P}\le M\rho^k.
\]
The first inequality in \eqref{eq:pc-simple-feedback-difference}
gives the stated bound on $K^{[k]}-K^*$, and
\eqref{eq:pc-simple-bounds} at $P^*$ gives
$\|F(P^*)\|_2\le\rho<1$.
If $s=0$, then $M=0$ and the feedback bound also vanishes,
so every $P^{[k]}$ and $K^{[k]}$ is $\mathbf 0$.
This proves item~3.

Every iterate lies in $\mathcal P_M$, thus $H(P)$ is
nonsingular, verifying Condition~\ref{assup:converge}(i).
The convergence of $P^{[k]}$ and $K^{[k]}$ established above
verifies Condition~\ref{assup:converge}(ii), while
$\|F(P^*)\|_2\le\rho<1$ verifies
Condition~\ref{assup:converge}(iii).
Thus Condition~\ref{assup:converge} holds, proving item~1.
\end{proof}

\section{Proof of Lemma~\ref{lm:convergence}}
\label{app:Proof of Lemma}

\begin{proof} 
Under Condition~\ref{assup:converge}(i), Lemma~\ref{lm:Uniqueness_FNE} guarantees a unique FNE for every finite-horizon game with zero reference and zero terminal cost. The backward recursion uses the same maps as the Riccati iteration starting from $P^{[0]}=\boldsymbol 0$. Thus, for $1\le t\le T$, the matrices $K_t^{i*}(T)$ and $P_t^{i*}(T)$ are the ith row blocks of $K^{[T-t]}$ and $P^{[T-t+1]}$, respectively. For every fixed $t\in\mathbb N_+$ and every $i\in\mathcal N$, Condition~\ref{assup:converge}(ii) therefore gives
\[
\lim_{T\to\infty}K_t^{i*}(T)=K^{i*},
\qquad
\lim_{T\to\infty}P_t^{i*}(T)=P^{i*}.
\]

By induction from $P^{[0]}=\boldsymbol 0$ and the definition of $\Phi$, every row block of $P^{[k]}$ is positive semidefinite. Hence $P^{i*}\succeq \boldsymbol 0$ for all $i\in\mathcal N$. Define
\[
F^*:=A+\sum_{j\in\mathcal N}B^jK^{j*},
\qquad
G^*:=C+\sum_{j\in\mathcal N}D^jK^{j*}.
\]
Taking limits in~\eqref{eq:K} and~\eqref{eq:P} along the Riccati iteration yields
\begin{equation}
\delta^i(B^i)^\top P^{i*}F^*+(D^i)^\top Q^iG^*+R^{ii}K^{i*}=0,
\label{eq:K'}
\end{equation}
and
\begin{equation}
P^{i*}=(G^*)^\top Q^iG^*+\delta^i(F^*)^\top P^{i*}F^*+\sum_{j\in\mathcal N}(K^{j*})^\top R^{ij}K^{j*}.
\label{eq:P'}
\end{equation}
Condition~\ref{assup:converge}(iii) gives $\|F^*\|_2<1$, so the limiting strategy profile is stabilizing.

We now show that no admissible unilateral deviation can reduce a player's cost. Fix $i\in\mathcal N$ and suppose that all other players use $u_t^j=K^{j*}x_t$. Consider any admissible deviation $u_t^i=\hat K^i x_t$, as defined in Section~\ref{sec:Infinite-Horizon Discrete-Time LQ}. The resulting closed-loop matrix
\[
A+B^i\hat K^i+\sum_{j\ne i}B^jK^{j*}
\]
is Schur stable. Thus, the induced cost is finite and $x_t\to \boldsymbol 0$ for every initial state.

Let
\[
e_t^i:=u_t^i-K^{i*}x_t,
\]
and
\[
M^i:=R^{ii}+(D^i)^\top Q^iD^i+\delta^i(B^i)^\top P^{i*}B^i.
\]
Since $R^{ii}\succ \boldsymbol 0$, $Q^i\succeq \boldsymbol 0$, and $P^{i*}\succeq \boldsymbol 0$, we have $M^i\succ \boldsymbol 0$. Along the deviating trajectory,
\[
x_{t+1}=F^*x_t+B^ie_t^i,
\qquad
y_t=G^*x_t+D^ie_t^i.
\]
Denote the undiscounted stage cost by 
\[
c_t^i:=\frac12\left[y_t^\top Q^iy_t+\sum_{j\in\mathcal N}(u_t^j)^\top R^{ij}u_t^j\right].
\]
Expanding these expressions and using~\eqref{eq:K'}--\eqref{eq:P'} gives
\[
c_t^i+\frac{\delta^i}{2}x_{t+1}^\top P^{i*}x_{t+1}-\frac12x_t^\top P^{i*}x_t
=\frac12(e_t^i)^\top M^ie_t^i.
\]
The quadratic state terms cancel by~\eqref{eq:P'}, and the terms involving both $x_t$ and $e_t^i$ cancel by~\eqref{eq:K'}.

Multiply this identity by $(\delta^i)^{t-1}$ and sum from $t=1$ to $L$. Since $x_t\to \boldsymbol 0$ and $0<\delta^i\le1$,
\[
\lim_{L\to\infty}(\delta^i)^Lx_{L+1}^\top P^{i*}x_{L+1}=0.
\]
Letting $L\to\infty$ yields
\[
J^i(x_1,u)=\frac12x_1^\top P^{i*}x_1+\frac12\sum_{t=1}^{\infty}(\delta^i)^{t-1}(e_t^i)^\top M^ie_t^i
\ge\frac12x_1^\top P^{i*}x_1.
\]
When player $i$ also uses $u_t^i=K^{i*}x_t$, we have $e_t^i= \boldsymbol 0$ for all $t$. This strategy is admissible because $F^*$ is Schur stable, and its cost is
\[
J^i(x_1)=\frac12x_1^\top P^{i*}x_1.
\]
Hence, no admissible unilateral deviation can reduce player $i$'s cost.

This argument holds for every player and every initial state. It also applies from any stage onward: the dynamics and stage costs are time invariant, and each player's continuation cost is scaled by a positive discount factor. Therefore, the limiting strategy profile constitutes an FNE. 
\end{proof}

\section{Proof of Theorem \ref{theo:finite-horizon strategy}}
\label{app:Proof of Theorem}
\begin{proof} 
Let 
\[
\tilde K^i:=K_1^{i*}(T^i),
\qquad
\tilde K:=\operatorname{col}
(\tilde K^1,\ldots,\tilde K^N),
\]
and define
\[
\tilde F:=A+B\tilde K,
\qquad
\tilde G:=C+D\tilde K,
\qquad
F^*:=A+BK^*,
\qquad
G^*:=C+DK^*.
\]
By the definition of $\epsilon$,
\[
\|\tilde K-K^*\|_{\max,K}=\epsilon,
\qquad
\|\tilde F-F^*\|_2\le b\epsilon,
\qquad
\|\tilde G-G^*\|_2\le d\epsilon.
\]
For a stationary feedback profile $K$, denote its
one-stage cost matrix for player $i$ by
\[
W^i(K):=(C+DK)^\top Q^i(C+DK)
+\sum_{j\in\mathcal N}(K^j)^\top R^{ij}K^j.
\]

\smallskip
\noindent\textbf{1. Stability and Cost Convergence.}
Condition~\ref{assup:converge} gives
$\|F^*\|_2<1$ and $\epsilon\to0$ as $T_h\to\infty$.
Fix $\mu\in(\|F^*\|_2,1)$.
There exists $T_0\in\mathbb N_+$ such that, whenever
$T_h\ge T_0$,
\[
\epsilon\le1,
\qquad
\|\tilde F\|_2
\le\|F^*\|_2+b\epsilon
\le\mu<1.
\]
Thus the resulting closed-loop system is stable.

Since $0<\delta^i\le1$, the series
\[
\tilde P^i
:=
\sum_{t=0}^{\infty}
(\delta^i)^t
(\tilde F^t)^\top
W^i(\tilde K)
\tilde F^t
\]
converges and defines a symmetric positive semidefinite
matrix. Substituting $x_t=\tilde F^{t-1}x_1$
into the cost function yields
\[
\tilde J^i(x_1;T^1,\ldots,T^N)
=\frac12x_1^\top\tilde P^i x_1.
\]
This representation determines $\tilde P^i$
uniquely among symmetric matrices.

The series above and the limiting Riccati equation give,
respectively,
\begin{align*}
\tilde P^i
=W^i(\tilde K)
+\delta^i\tilde F^\top
\tilde P^i\tilde F, \quad
P^{i*}
=W^i(K^*)
+\delta^i(F^*)^\top P^{i*}F^*.
\end{align*}
We compare these equations using the elementary inequality
\begin{equation}
\|X^\top SX-Y^\top SY\|_2
\le\|S\|_2(\|X\|_2+\|Y\|_2)\|X-Y\|_2.
\label{eq:quadratic-difference}
\end{equation}
Indeed, $\epsilon\le1$ implies
\[
\|\tilde G\|_2\le\|G^*\|_2+d,
\qquad
\|\tilde K\|_{\max,K}\le\|K^*\|_{\max,K}+1.
\]
Applying the preceding inequality to the output and
control terms gives
\[
\begin{aligned}
\|W^i(\tilde K)-W^i(K^*)\|_2
&\le q(\|\tilde G\|_2+\|G^*\|_2)
\|\tilde G-G^*\|_2\\
&\quad+\sum_{j\in\mathcal N}\|R^{ij}\|_2
(\|\tilde K^j\|_2+\|K^{j*}\|_2)
\|\tilde K^j-K^{j*}\|_2\\
&\le qd(2\|G^*\|_2+d\epsilon)\epsilon
+\overline R(2\|K^*\|_{\max,K}+\epsilon)\epsilon\\
&\le
\bigl[
qd(2\|G^*\|_2+d)
+\overline R(2\|K^*\|_{\max,K}+1)
\bigr]\epsilon.
\end{aligned}
\]
Here we used
$\|\tilde G\|_2\le\|G^*\|_2+d\epsilon$,
$\|\tilde K^j\|_2\le\|K^{j*}\|_2+\epsilon$,
and $\epsilon\le1$. 
Similarly, since $\|\tilde F\|_2,\|F^*\|_2<1$,
\[
\|\tilde F^\top P^{i*}\tilde F
-(F^*)^\top P^{i*}F^*\|_2
\le2b\|P^*\|_{\max,P}\epsilon.
\]
Subtracting the two cost equations, and adding and
subtracting
$\delta^i\tilde F^\top P^{i*}\tilde F$,
now yields
\[
\begin{aligned}
\|\tilde P^i-P^{i*}\|_2
\le
\delta^i\|\tilde F\|_2^2
\|\tilde P^i-P^{i*}\|_2+L\epsilon \le
\mu^2\|\tilde P^i-P^{i*}\|_2+L\epsilon,
\end{aligned}
\]
where the last inequality uses
$0<\delta^i\le1$ and $\|\tilde F\|_2\le\mu$, and 
\[
L:=
qd(2\|G^*\|_2+d)
+\overline R(2\|K^*\|_{\max,K}+1)
+2b\|P^*\|_{\max,P}.
\]
Consequently, with $c:=(L+1)/(1-\mu^2)>0$,
\[
\max_{i\in\mathcal N}
\|\tilde P^i-P^{i*}\|_2
\le c\epsilon.
\]
The constants $T_0$ and $c$ are independent of the
individual horizons and the initial state.

Finally,
\[
\left|
\tilde J^i(x_1;T^1,\ldots,T^N)-J^i(x_1)
\right|
=
\frac12
\left|x_1^\top(\tilde P^i-P^{i*})x_1\right|
\le
\frac c2\|x_1\|_2^2\epsilon.
\]
Since $\epsilon\to0$ as $T_h\to\infty$, this proves
all statements in part~1.

\smallskip
\noindent\textbf{2. Explicit Convergence Bounds.}

Suppose that the additional parameter conditions
in Theorem~\ref{theo:Sufficient parameter conditions} hold.
By \eqref{eq:pc-simple-bounds},
\[
\|\mathcal K(P)\|_{\max,K}
\le\frac12\sqrt{\frac{s}{\overline R}},
\qquad P\in\mathcal P_M.
\] 
Every Riccati iterate and the limit $P^*$ belong to
$\mathcal P_M$.
Moreover, $\tilde K^i$ is the $i$th row block of
$\mathcal K(P^{[T^i-1]})$. For every $i\in\mathcal N$, since
$P^{[T^i-1]}\in\mathcal P_M$, we have
\[
\begin{aligned}
\|\tilde K^i\|_2
 =\|\mathcal K^i(P^{[T^i-1]})\|_2 \le\|\mathcal K(P^{[T^i-1]})\|_{\max,K} \le\frac12\sqrt{\frac{s}{\overline R}}, \ \|K^*\|_{\max,K} \le\frac12\sqrt{\frac{s}{\overline R}}.
\end{aligned}
\]
Using the parameter conditions, we obtain
\[
\|\tilde F\|_2,
\ \|F^*\|_2
\le
a+\frac b2\sqrt{\frac{s}{\overline R}}
\le\rho<1,
\]
and
\[
\|\tilde G\|_2,
\ \|G^*\|_2
\le
\|C\|_2+\frac d2\sqrt{\frac{s}{\overline R}}
\le\frac{11}{10}\|C\|_2.
\]
In particular, stability holds for every choice
$T^1,\ldots,T^N\in\mathbb N_+$.
Thus the cost matrices and their equations in part~1
are well defined for all such horizons.

The first-stage feedback and cost matrices correspond
to the $i$th blocks of $K^{[T^i-1]}$ and $P^{[T^i]}$,
respectively.
Theorem~\ref{theo:Sufficient parameter conditions} therefore gives
\[
\|K_1^{i*}(T^i)-K^{i*}\|_2
\le\frac{5bM}{4r}\rho^{T^i},
\qquad
\|P_1^{i*}(T^i)-P^{i*}\|_2
\le M\rho^{T^i}.
\]
In particular,
\[
\epsilon
\le\frac{5bM}{4r}\rho^{T_h}.
\]

It remains to bound the resulting infinite-horizon
cost matrices. 
Subtracting the two cost equations and adding and
subtracting $\delta^i\tilde F^\top P^{i*}\tilde F$ gives
\[
\begin{aligned}
\tilde P^i-P^{i*}
=W^i(\tilde K)-W^i(K^*)  +\delta^i\tilde F^\top
(\tilde P^i-P^{i*})\tilde F  +\delta^i\bigl[
\tilde F^\top P^{i*}\tilde F
-(F^*)^\top P^{i*}F^*
\bigr].
\end{aligned}
\]
Taking norms, we obtain
\[
\begin{aligned}
\|\tilde P^i-P^{i*}\|_2
 \le\|W^i(\tilde K)-W^i(K^*)\|_2 +\delta^i\|\tilde F\|_2^2
\|\tilde P^i-P^{i*}\|_2 +\delta^i
\|\tilde F^\top P^{i*}\tilde F
-(F^*)^\top P^{i*}F^*\|_2.
\end{aligned}
\]
By \eqref{eq:quadratic-difference} and the bounds
$\|\tilde G-G^*\|_2\le d\epsilon$ and
$\|\tilde K-K^*\|_{\max,K}=\epsilon$,
\[
\begin{aligned}
\|W^i(\tilde K)-W^i(K^*)\|_2
&\le\bigl[
qd(\|\tilde G\|_2+\|G^*\|_2) +\overline R
(\|\tilde K\|_{\max,K}+\|K^*\|_{\max,K})
\bigr]\epsilon\\
&\le\left(
\frac{11}{5}q\|C\|_2d+\sqrt{s\overline R}
\right)\epsilon.
\end{aligned}
\]
Similarly, using $\|P^{i*}\|_2\le M$ and
$\|\tilde F-F^*\|_2\le b\epsilon$ gives
\[
\begin{aligned}
\|\tilde F^\top P^{i*}\tilde F
-(F^*)^\top P^{i*}F^*\|_2
&\le Mb(\|\tilde F\|_2+\|F^*\|_2)\epsilon \le2M\rho b\epsilon.
\end{aligned}
\]
Substituting these estimates and using
$0<\delta^i\le1$ and $\|\tilde F\|_2\le\rho$, we obtain
\[
\|\tilde P^i-P^{i*}\|_2
\le\rho^2\|\tilde P^i-P^{i*}\|_2
+\left(
\frac{11}{5}q\|C\|_2d
+\sqrt{s\overline R}+2M\rho b
\right)\epsilon.
\] 
Rearranging yields
\[
(1-\rho^2)\|\tilde P^i-P^{i*}\|_2
\le
\left(
\frac{11}{5}q\|C\|_2d
+\sqrt{s\overline R}+2M\rho b
\right)\epsilon.
\]
The parameter conditions imply
\[
q\|C\|_2d\le\frac15\sqrt{s\overline R},
\qquad
M\rho b\le\frac25\sqrt{s\overline R}.
\]
Consequently,
\[
\max_{i\in\mathcal N}
\|\tilde P^i-P^{i*}\|_2
\le
\frac{3\sqrt{s\overline R}}{1-\rho^2}\epsilon.
\]
Combining the bound on $\epsilon$ with
\[
\frac{b\sqrt{s\overline R}}{r}
=\kappa\|C\|_2b
\le\frac{1-a}{5}
\]
yields
\[
\max_{i\in\mathcal N}
\|\tilde P^i-P^{i*}\|_2
\le
\frac{3(1-a)}{4(1-\rho^2)}
M\rho^{T_h}.
\]
By the definition $\rho=a+(1-a)/10$,
\[
\frac{3(1-a)}{4(1-\rho^2)}
=
\frac{5}{6(1+\rho)}
\le1.
\]
Therefore,
\[
\max_{i\in\mathcal N}
\|\tilde P^i-P^{i*}\|_2
\le M\rho^{T_h}.
\]
The corresponding cost bound follows immediately:
\[
\left|
\tilde J^i(x_1;T^1,\ldots,T^N)-J^i(x_1)
\right|
\le
\frac12\|x_1\|_2^2
\|\tilde P^i-P^{i*}\|_2
\le
\frac M2\|x_1\|_2^2\rho^{T_h}.
\]
This completes the proof.
\end{proof}

\section{Infinitely Many FNEs in Numerical Example 2}
\label{app:simulation-fnes}

The construction starts with two stable solutions of a scalar game.
We place these solutions in two independent directions and rotate
them to obtain infinitely many planar feedbacks. We then add the remaining state coordinate and the fixed
nonzero references to recover Example 2. 

\smallskip
\noindent\textbf{1. Two Scalar Solutions.}
Consider a scalar state $z$ and two scalar controls $\alpha,\beta$:
\[
z^+=a_0z+b_0(\alpha+\beta),\qquad
y=c_0z+d_0(\alpha+\beta).
\]
The two players have stage costs
\[
\frac12\bigl[y^2+r\alpha^2+r_{\mathrm c}\beta^2\bigr],
\qquad
\frac12\bigl[y^2+r_{\mathrm c}\alpha^2+r\beta^2\bigr],
\]
and the same discount factor $\delta$. We use the coefficients
from Example 2:
$a_0=0.95$, $b_0=0.01$, $c_0=0.0196$, $d_0=-0.002$,
$r=0.1$, $r_{\mathrm c}=0.5$, and $\delta=0.99$.

Seek symmetric feedbacks $\alpha=\beta=kz$ with a common value
function $pz^2/2$. Fix the second player's feedback at $\beta=kz$.
With the candidate value function $pz^2/2$, the first player's
Bellman equation is
\[
\begin{aligned}
\frac12pz^2=\min_{\alpha\in\mathbb R}\Bigl\{
&\frac12\bigl[(c_0z+d_0(\alpha+kz))^2
+r\alpha^2+r_{\mathrm c}k^2z^2\bigr] +\frac{\delta p}{2}
\bigl[a_0z+b_0(\alpha+kz)\bigr]^2
\Bigr\}.
\end{aligned}
\]
Differentiating the expression on the right with respect to
$\alpha$ gives the first-order condition
\[
0=r\alpha+d_0[c_0z+d_0(\alpha+kz)]
+\delta b_0p[a_0z+b_0(\alpha+kz)].
\]
At a symmetric solution, $\alpha=kz$.
Substituting this into the first-order condition and the
Bellman equation, and matching the coefficients of $z$
and $z^2$, respectively, gives
\begin{equation}
\begin{aligned}
0&=rk+d_0(c_0+2d_0k)
+\delta b_0p(a_0+2b_0k),\\
p&=(c_0+2d_0k)^2+(r+r_{\mathrm c})k^2
+\delta(a_0+2b_0k)^2p.
\end{aligned}
\label{eq:scalar-feedback-value}
\end{equation}
The second player gives the same equations by symmetry.
For $p\ge0$, the first equation gives
\[
m(p):=r+2d_0^2+2\delta b_0^2p>0,\qquad
k(p):=-\frac{d_0c_0+\delta b_0a_0p}{m(p)}.
\]
Set $$f(p):=a_0+2b_0k(p)$$ for the closed-loop multiplier.
Substitute $k=k(p)$ into the second equation
of~\eqref{eq:scalar-feedback-value}. 
The left-hand side minus the right-hand side is continuous
and has opposite signs at the endpoints of both
$(0.003,0.004)$ and $(219,221)$.
Thus there are roots
\[
p_L\in(0.003,0.004),\qquad p_H\in(219,221).
\]
Both satisfy $0<f(p)<1$. These are the two stable scalar
solutions used below.

\smallskip
\noindent\textbf{2. Two Dimensions and Rotation.}
Place the two scalar solutions along two orthogonal directions.
For the planar rotation $U_\theta$, $0\le\theta\le\pi/2$, set
\[
\Pi_{\mathrm p}
=U_\theta\operatorname{diag}(p_L,p_H)U_\theta^\top,
\qquad
K_{\mathrm p}
=U_\theta\operatorname{diag}\bigl(k(p_L),k(p_H)\bigr)U_\theta^\top.
\]
The subscript $\mathrm p$ denotes the planar part. Define
\[
F_{\mathrm p}=a_0I_2+2b_0K_{\mathrm p},
\qquad
G_{\mathrm p}=c_0I_2+2d_0K_{\mathrm p}.
\]
Apply~\eqref{eq:scalar-feedback-value} first with
$p=p_L$, $k=k(p_L)$, and then with $p=p_H$, $k=k(p_H)$.
Place the two scalar identities on the diagonal of
$2\times2$ matrices, and multiply on the left by $U_\theta$
and on the right by $U_\theta^\top$.
Using $U_\theta^\top U_\theta=I_2$ and the definitions of
$\Pi_{\mathrm p},K_{\mathrm p},F_{\mathrm p},G_{\mathrm p}$,
we obtain
\[
\begin{aligned}
rK_{\mathrm p}+d_0G_{\mathrm p}
+\delta b_0\Pi_{\mathrm p}F_{\mathrm p}
=\mathbf0, \quad 
\Pi_{\mathrm p}
=G_{\mathrm p}^\top G_{\mathrm p}
+(r+r_{\mathrm c})K_{\mathrm p}^\top K_{\mathrm p}
+\delta F_{\mathrm p}^\top\Pi_{\mathrm p}F_{\mathrm p}.
\end{aligned}
\]
These identities hold for every $\theta$.
Moreover, $k'(p)<0$, so
\[
(K_{\mathrm p})_{11}
=k(p_L)\cos^2\theta+k(p_H)\sin^2\theta
\]
is strictly decreasing on $[0,\pi/2]$. 
Thus, varying the rotation angle $\theta$ gives infinitely many
distinct feedback matrices $K_{\mathrm p}$.
Each $K_{\mathrm p}$, together with its corresponding
$\Pi_{\mathrm p}$, satisfies the feedback and value equations
for the two-dimensional state.

We next add the third state coordinate and the fixed nonzero
references. We will show that every rotated pair can be used
to construct a stabilizing affine FNE of the original game
with four players and three state coordinates.

\smallskip
\noindent\textbf{3. Adding the Third State.}
The matrices in Example 2 have the form
\[
\begin{aligned}
A&=\begin{bmatrix}a_0I_2&v_A\\\mathbf0^\top&h\end{bmatrix},
& B&=b_0[E\ EU],\\
C&=[c_0I_2\ v_C],
& D&=d_0[I_2\ U],
\end{aligned}
\]
where $E=[I_2\ \mathbf0]^\top$, $h=0.75$,
$v_A=(0.01,0.01)^\top$, $v_C=(0.005,0.005)^\top$, and
$U=\left[\begin{smallmatrix}0.8&-0.6\\0.6&0.8\end{smallmatrix}\right]$.
Also, $Q^i=I_2$ and $\delta^i=\delta$ for every player.

For each rotated pair $(K_{\mathrm p},\Pi_{\mathrm p})$,
we will construct matrices $K_\theta,\Pi_\theta$ satisfying
the feedback and quadratic value equations for the full
three-state system.

\smallskip
\noindent\textbf{3.1. Extending the Feedback.}
We retain $K_{\mathrm p}$ and add a column for the third state:
\[
K_{\mathrm g}=[K_{\mathrm p}\ \xi],\qquad
K_\theta=\operatorname{col}(K_{\mathrm g},U^\top K_{\mathrm g}),
\]
where $\xi\in\mathbb R^2$ will be determined in Step~3.3.
For any choice of $\xi$,
\[
F_\theta=A+BK_\theta
=\begin{bmatrix}
F_{\mathrm p}&v_A+2b_0\xi\\
\mathbf0^\top&h
\end{bmatrix}.
\]
Recall that $f(p)=a_0+2b_0k(p)$ and
\[
F_{\mathrm p}
=U_\theta\operatorname{diag}(f(p_L),f(p_H))U_\theta^\top.
\]
Since $F_\theta$ is block upper triangular, its eigenvalues
are $f(p_L),f(p_H),h$, all in $(0,1)$.
Thus $F_\theta$ is Schur stable for every $\xi$.

\smallskip
\noindent\textbf{3.2. Finding the Cost Matrix.}
Fix $\xi$ temporarily and consider the cost under
$u=K_\theta x$. Set
\[
G_\theta=C+DK_\theta
=[G_{\mathrm p}\ \ v_C+2d_0\xi].
\]
The two control pairs are
\[
\begin{pmatrix}u^1\\u^2\end{pmatrix}=K_{\mathrm g}x,
\qquad
\begin{pmatrix}u^3\\u^4\end{pmatrix}=U^\top K_{\mathrm g}x.
\]
Orthogonality of $U$ gives
$\|U^\top K_{\mathrm g}x\|_2^2=\|K_{\mathrm g}x\|_2^2$.
Since
\[
(R^{i1},R^{i2},R^{i3},R^{i4})
=
\begin{cases}
(r,r,r_{\mathrm c},r_{\mathrm c}), & i=1,2,\\
(r_{\mathrm c},r_{\mathrm c},r,r), & i=3,4,
\end{cases}
\]
every player has the same quadratic control-cost matrix:
\[
\sum_{j=1}^4(K_\theta^j)^\top R^{ij}K_\theta^j
=(r+r_{\mathrm c})K_{\mathrm g}^\top K_{\mathrm g},
\qquad i=1,\ldots,4.
\]

Let $V^i(x)$ denote player $i$'s total discounted cost under
this fixed feedback. The next state is $F_\theta x$, so
\[
\begin{aligned}
V^i(x)
=\frac12(G_\theta x-l^i)^\top Q^i(G_\theta x-l^i)  +\frac12\sum_{j=1}^4
(K_\theta^j x)^\top R^{ij}(K_\theta^j x)
+\delta^i V^i(F_\theta x).
\end{aligned}
\]
Write the quadratic part of $V^i(x)$ as
$\tfrac12x^\top\Pi_\theta^i x$.
Comparing quadratic terms gives
\[
\begin{aligned}
x^\top\Pi_\theta^i x
=x^\top G_\theta^\top Q^iG_\theta x
+\sum_{j=1}^4
x^\top(K_\theta^j)^\top R^{ij}K_\theta^j x +\delta^i x^\top
F_\theta^\top\Pi_\theta^i F_\theta x.
\end{aligned}
\]
This holds for every $x$. Using $Q^i=I_2$,
$\delta^i=\delta$, and the control-cost identity above,
we obtain
\begin{equation}
\Pi_\theta^i
=G_\theta^\top G_\theta
+(r+r_{\mathrm c})K_{\mathrm g}^\top K_{\mathrm g}
+\delta F_\theta^\top\Pi_\theta^i F_\theta.
\label{eq:third-state-value}
\end{equation}
Since $F_\theta$ is Schur stable, this equation has a unique
positive semidefinite solution. Thus all four players have
the same quadratic value matrix, denoted by $\Pi_\theta$.
This determines $\Pi_\theta$ for each choice of $\xi$.

\smallskip
\noindent\textbf{3.3. Determining the Remaining Feedback Column.}
We now choose $\xi$ so that
\begin{equation}
rK_\theta+D^\top G_\theta
+\delta B^\top\Pi_\theta F_\theta=\mathbf0.
\label{eq:third-state-feedback}
\end{equation}
Since the third row of $B$ is zero, this equation involves
only the first two rows of $\Pi_\theta$.
Its upper-left $2\times2$ block satisfies the planar value
equation in Step~2, so it equals $\Pi_{\mathrm p}$ by uniqueness.
Denote its upper-right block by $\zeta\in\mathbb R^2$.

The first two columns of~\eqref{eq:third-state-feedback}
already hold by Step~2.
The equations for players $3,4$ are $U^\top$ times those
for players $1,2$, so its third column reduces to
\[
r\xi+d_0(v_C+2d_0\xi)
+\delta b_0\bigl[\Pi_{\mathrm p}(v_A+2b_0\xi)+h\zeta\bigr]
=\mathbf0.
\]
Meanwhile, the upper-right block of
\eqref{eq:third-state-value} gives
\[
\begin{aligned}
\zeta
&=G_{\mathrm p}^\top(v_C+2d_0\xi)
+(r+r_{\mathrm c})K_{\mathrm p}^\top\xi +\delta F_{\mathrm p}^\top
\bigl[\Pi_{\mathrm p}(v_A+2b_0\xi)+h\zeta\bigr].
\end{aligned}
\]
The planar matrices are symmetric, and Step~2 gives
$d_0G_{\mathrm p}+\delta b_0\Pi_{\mathrm p}F_{\mathrm p}
=-rK_{\mathrm p}$.
Using the transpose of this identity and collecting terms,
the two equations become
\[
\begin{aligned}
H_{\mathrm p}\xi+\delta b_0h\zeta
&=-d_0v_C-\delta b_0\Pi_{\mathrm p}v_A,\\
(I_2-\delta hF_{\mathrm p})\zeta
&=G_{\mathrm p}v_C+\delta F_{\mathrm p}\Pi_{\mathrm p}v_A
+(r_{\mathrm c}-r)K_{\mathrm p}\xi,
\end{aligned}
\]
where $H_{\mathrm p}=(r+2d_0^2)I_2+2\delta b_0^2\Pi_{\mathrm p}$.

These equations are linear in $\xi,\zeta$.
In the basis given by $U_\theta$, they split into two
$2\times2$ systems with determinants
\[
e_h(p)=m(p)(1-\delta hf(p))
+\delta b_0h(r_{\mathrm c}-r)k(p),
\qquad p\in\{p_L,p_H\},
\]
where $m(p)=r+2d_0^2+2\delta b_0^2p$.
The root intervals give $e_h(p)>0.02$ for both roots.
Hence $\xi,\zeta$ are uniquely determined for every $\theta$.
Since $I_2-\delta hF_{\mathrm p}$ is invertible, the second
equation ensures that the computed $\zeta$ is the value
cross block corresponding to this $\xi$.
Thus~\eqref{eq:third-state-feedback} holds.

We have now constructed the feedback and quadratic value
matrices for all three state coordinates.
We next determine the constant control terms for the fixed
references.

\smallskip
\noindent\textbf{4. Adding the References and Verifying Equilibrium.}
The fixed references affect the linear and constant value terms.
Write
\[
u=K_\theta x+L,\qquad
V^i(x)=\tfrac12x^\top\Pi_\theta x+(S^i)^\top x+w^i,
\]
and let $R^i=\operatorname{diag}(R^{i1},\ldots,R^{i4})$.
Using the feedback identity above, the remaining linear value
equations and first-order conditions become
\[
\begin{aligned}
&(I_3-\delta F_\theta^\top)S^i
=K_\theta^\top(R^i-rI_4)L-G_\theta^\top l^i,\\
&(rI_4+D^\top D+\delta B^\top\Pi_\theta B)_{i,:}L
+\delta(B^i)^\top S^i
=(D^i)^\top l^i.
\end{aligned}
\]
Since $F_\theta$ is Schur stable, the first equation determines
$S^i$ from $L$.
Substitute it into the second equation.
Because $F_\theta$ is block upper triangular and the third row
of $B$ is zero, only the planar blocks enter the coefficient matrix.

To solve this system, align the two control pairs, take their
sum and difference, and rotate to the planar eigenbasis:
\[
U_\theta^\top
\left[
\begin{pmatrix}L^1\\L^2\end{pmatrix}
\pm U\begin{pmatrix}L^3\\L^4\end{pmatrix}
\right].
\]
In these coordinates, the coefficient matrix is diagonal.
Its four entries are $\lambda_\pm(p_L),\lambda_\pm(p_H)$, where
\[
\lambda_+(p)=m(p)+
\frac{\delta b_0(r_{\mathrm c}-r)k(p)}{1-\delta f(p)},
\qquad
\lambda_-(p)=r-
\frac{\delta b_0(r_{\mathrm c}-r)k(p)}{1-\delta f(p)}.
\]
The root intervals give
\[
\lambda_+(p_L)>0.1,\qquad
\lambda_+(p_H)<-0.01,\qquad
\lambda_-(p_L),\lambda_-(p_H)>0.09.
\]
Thus $L$ and every $S^i$ exist uniquely for the fixed references.
The constant value equations determine $w^i$ uniquely because
$1-\delta>0$.

It remains to check that these solutions are equilibria.
Completing the square gives
\[
\begin{aligned}
&c^i(x,u^i,u^{-i}_\theta)+\delta V^i(x^+)-V^i(x)\\
&\quad=\tfrac12\bigl[r+(D^i)^\top D^i
+\delta(B^i)^\top\Pi_\theta B^i\bigr]
(u^i-K_\theta^ix-L^i)^2\ge0.
\end{aligned}
\]
Here $c^i$ is the stage cost, $x^+=Ax+Bu$, and the other
players use their constructed affine controls.
Under any admissible stationary affine unilateral deviation,
the state is bounded and the discounted terminal value vanishes.
Summing with discount weights proves the Nash inequality for
every state and starting stage, with equality under the
constructed strategy.

The extensions in Steps~3--4 leave $K_{\mathrm p}$ unchanged.
Hence the resulting FNEs are distinct by Step~2.
Their common value matrices satisfy $\|\Pi_\theta\|_2\ge p_H>219$.
On the other hand,
\[
\|A\|_2^2\le\|A\|_1\|A\|_\infty=0.912,\qquad
\|C\|_2^2=c_0^2+\|v_C\|_2^2<4.4\times10^{-4},
\]
so $M=2\|C\|_2^2/(1-\|A\|_2^2)<0.01$.
Their stacked value matrices therefore lie outside
$\mathcal P_M$, where the Riccati map is a contraction.

\end{document}